\documentclass[fleqn,usenatbib]{mnras}

\usepackage{newtxtext,newtxmath}
\usepackage{amsmath,mathtools,xfrac}
\usepackage{graphicx}

\usepackage[T1]{fontenc}

\DeclareRobustCommand{\VAN}[3]{#2}
\let\VANthebibliography\thebibliography
\def\thebibliography{\DeclareRobustCommand{\VAN}[3]{##3}\VANthebibliography}

\usepackage{graphicx}	
\usepackage{amsmath}	

\title[Cross correlation at high spectral resolution]{Revising the cross correlation technique at high spectral resolution}

\author[Zamora \& Díaz]{
S. Zamora,$^{1,2}$\thanks{E-mail: sandra.zamora@uam.es}
A. I. Díaz$^{1,2}$
\\
$^{1}$Departamento de Física Teórica, Universidad Autónoma de Madrid, 28049 Madrid, Spain\\
$^{2}$CIAFF, Universidad Autónoma de Madrid, 28049 Madrid, Spain\\
}

\date{Accepted XXX. Received YYY; in original form ZZZ}

\pubyear{2015}

\begin{document}
\label{firstpage}
\pagerange{\pageref{firstpage}--\pageref{lastpage}}
\maketitle

\begin{abstract}

Cross-correlation techniques have been used since 1974 for measuring velocity shifts and velocity dispersions from stellar and nebular spectra and, since 1979, the analysis based on the Fourier Method has been applied. However, we are currently obtaining data with spectral resolutions higher than those for which this technique was developed, hence some revision seems timely. The principal aim of this work is adapting Tonry and Davis' method and implementing it for the treatment of very high spectral resolution data. We have applied this technique to two different sets of spectroscopic data of moderate and high resolutions obtained with the MUSE and MEGARA spectrographs respectively. Using stellar spectra obtained with these two instruments (i) we have optimised the input parameters; (ii) we have analysed the method assumptions; and (iii) we have compared the results for the two sets of data. For MEGARA data, we have found that the cross-correlation function lost its Gaussian behavior at higher resolutions. Thus, we have developed an equivalent mathematical method that can be used for this kind of data. Additionally, the velocity dispersion error analysis suggests that the greatest error introduced in this method is due to the subtraction or masking of the nebular lines. For the application of cross-correlation techniques to high spectral resolution data, we propose to calculate the galaxy-galaxy and star-galaxy correlations, with widths $\mu_{gg}$ and $\mu_{gt}$ respectively. Then, the width of the broadening function can be calculated as $\sigma = \sqrt{\mu_{gg}^2 - \mu_{gt}^2}$. 

\end{abstract}

\begin{keywords}
galaxies: star clusters: general -- galaxies: starburst -- techniques: imaging spectroscopy -- Astronomical instrumentation -- methods -- techniques
\end{keywords}


\section{Introduction}
Cross-correlation techniques have been applied since several decades ago \citet{1974A&A....31..129S} in order to derive objects’ redshifts and velocity dispersions by correlating the object's spectrum with that of a reference star and measuring respectively the peak center and width of the found cross-correlation function. In \citet{1979AJ.....84.1511T} a detailed correlation analysis based on the Fourier Method was presented. The use of the  Fourier Transform algorithm reduces the operational and computing times, providing a straightforward determination of the cross correlation function. Additionally, an internal error measure of the correlation peak can be directly performed. 

This technique can be applied to a large number of astronomical problems. Examples of these are: the derivation of the stellar velocity rotation, the gas and star  velocity dispersion in different types of galaxies (including broad and narrow line active galaxies), or the study of regions with active star formation  \citep[i.e.][]{2007MNRAS.378..163H,1995ApJS...99...67N}.

Over at least half-a-century, numerous thorough method reviews have been presented by various authors, beginning with \citet{1978ApJ...221....1D} and followed by \citet{2001BSAO...51...11M} or \citet{2006ApJ...641..117G}. However, we are currently obtaining data with instrument  of characteristics very different  from those used up to now that include a good number of spectrographs of very high spectral resolution. At this high resolutions, the cross correlation technique assumptions may not be correct, hence a revision of the method is required. 


In this work, we have tested the widely used Tonry and Davis' technique \citep{1979AJ.....84.1511T} using data of moderate and high spectral resolutions obtained with two spectrographs currently in use: MUSE (R ~ 2500) and MEGARA (R ~6000 - 20000) adapting this method to the analysis of high spectral resolution data. We have chosen these two instruments just for illustration purposes although the results obtained can be generalised to any other since they do not depend on a given spectrograph but only on the spectral resolution.  

In Section 2, we describe the observations used and the spectral resolution expected for each set of the data used. The cross correlation technique is introduced in Section 3 including the necessary mathematical concepts, the presentation of the original method, and the developing of the one proposed in this work, as well as the steps followed in the spectrum preparation required prior to the  application of the method. Our analyses and results are presented in Section 4 for the MUSE and MEGARA data respectively. A discussion of the method assumptions at different spectral resolutions, the comparison between the traditional methodology and the new one presented here, and the analysis of the different errors involved in them are given in Section 5. Finally, the summary and conclusions of this work are given in Section 6.

\section{Observations}
\label{sec:observations}
In this work, we have used data from two different instruments in order to compare the traditional Tonry and Davis' method \citep{1979AJ.....84.1511T} and the proposed one at high spectral resolutions. We have used data from the Multi-Unit Spectroscopic Explorer \citep[MUSE][]{MUSE} and from the Multi-Espectrógrafo en GTC de Alta Resolución para Astronomía \citep[MEGARA][]{2018SPIE10702E..16C}. Both instruments are integral-field spectrographs (IFS) and are attached to one of the Very Large Telescopes (VLT) of the European Southern Observatory (ESO, Chile) and at the 10.4-m GTC telescope at La Palma Observatory (Spain) respectively. MUSE has a nominal dispersion of 1.25 \AA /pixel with a spectral resolving power, R$_{FWHM}$ = $\lambda / \Delta \lambda$, from 1770 (at 4800 \AA ) to 3590 (at 9300 \AA ) in the blue and red arms respectively. It covers the visible wavelength range from 4800 \AA\ to 9300 \AA\ . On the other hand, MEGARA has three observing modes: low resolution LR, medium resolution MR and high resolution HR providing R values from $\sim$ 6000, 12000 and 20000 respectively, covering from 3654 \AA\ to 9635 \AA\ in different dispersion elements.

\begin{table*}
\centering
\caption{Selected giants and super-giants MUSE stars from \citet{2019A&A...629A.100I}.}
\label{tab:MUSE_stars}
\begin{tabular}{ccccccc}
\hline
Name     & RA         & DEC         & Sp Type     & Teff & Logg & Fe/H \\
\hline
HD   173158       & 18:43:45.3048425424 & +05:44:14.622467928 & K0      & 5164 $\pm$ 121 & 0.87 $\pm$ 0.43 & 0.04 $\pm$ 0.20  \\
HD   170820       & 18:32:13.1080081656 & -19:07:26.333290848 & K0III   & 4707 $\pm$ 57  & 1.65 $\pm$ 0.13 & 0.17             \\
HD   099998       & 11:30:18.8933524920 & -03:00:12.598924644 & K3.5III & 4001 $\pm$ 32  & 1.56 $\pm$ 0.20 & -0.24 $\pm$ 0.07 \\
HD   232078       & 19:38:12.0696686280 & +16:48:25.633514592 & K3IIp   & 4295 $\pm$ 48  & 0.82 $\pm$ 0.27 & -1.08 $\pm$ 0.11 \\
HD   114960       & 13:13:57.5656668240 & +01:27:23.203612764 & K5III   & 4000           &                 &                  \\
HD   306799       & 11:36:34.8400229976 & -61:36:35.189793504 & M0Iab   & 3650           &                 &                  \\
HD   102212       & 11:45:51.5595736    & +06:31:45.741287    & M1III   & 3738 $\pm$ 6   & 1.55 $\pm$ 0.10 & -0.41 $\pm$ 0.05 \\
HD   101712       & 11:41:49.4048035632 & -63:24:52.454152980 & M3Iab   & 3200           &                 &                  \\
HD   100733       & 11:35:13.2823419624 & -47:22:21.284741784 & M3III   & 3530           &                 &                  \\
IRAS   15060+0947 & 15:08:25.7458537200 & +09:36:18.371710980 & M9III   & 3281           &                 &                 \\
\hline
\end{tabular}
\end{table*}
\begin{table*}
\centering
\caption{Selected giants and super-giants MEGARA stars from \citet{2020MNRAS.493..871G}.}
\label{tab:MEGARA_stars}
\begin{tabular}{ccccccc}
\hline
Name     & RA J2000 (deg)         & DEC J2000 (deg)        & Spectral type     & Effective Temperature (K) & log(g) & {[}Fe/H{]} \\
\hline
HD042983 & 06:13:52.8 & +02:48:31.9 & K0          & 4633 & 3.23 & 0.00       \\
HD105087 & 12:06:00.9 & +14:38:56.7 & K0          & 4966 & 4.16 & -0.43      \\
HD215704 & 22:46:20.4 & +50:12:35.9 & K0          & 5418 & 4.20 & 0.07       \\
HD040801 & 06:03:17.9 & +42:54:41.5 & K01II       & 4740 & 2.94 & -0.04      \\
HD045829 & 06:30:02.3 & +07:55:16.0 & K0Iab       & 4500 & 0.20 & 0.00       \\
HD027371 & 04:19:47.6 & +15:37:39.5 & K0III       & 4956 & 2.71 & 0.07       \\
HD045410 & 06:30:47.1 & +58:09:45.5 & K0III       & 4838 & 2.34 & 0.17       \\
HD100696 & 11:36:02.8 & +69:19:22.6 & K0III       & 4890 & 2.27 & -0.25      \\
HD107328 & 12:20:21.0 & +03:18:45.3 & K0III       & 4380 & 2.39 & -0.48      \\
HD131111 & 14:50:29.6 & +37:16:19.4 & K0III       & 4710 & 3.11 & -0.29      \\
HD147677 & 16:22:05.8 & +30:53:31.2 & K0III       & 4910 & 2.98 & -0.08      \\
HD063302 & 07:47:38.5 & -15:59:26.5 & K2Iab       & 4500 & 0.20 & 0.12       \\
HD055280 & 07:15:54.9 & +59:38:14.9 & K2III       & 4623 & 2.54 & 0.08       \\
HD072184 & 08:32:55.0 & +38:00:58.9 & K2III       & 4624 & 2.61 & 0.12       \\
HD115136 & 13:13:28.0 & +67:17:16.6 & K2III       & 4541 & 2.40 & 0.05       \\
HD017506 & 02:50:41.8 & +55:53:43.8 & K3I         & 3500 & 1.00 & 0.09       \\
HD083425 & 09:38:27.3 & +04:38:57.4 & K3III       & 4120 & 1.77 & -0.29      \\
HD125560 & 14:19:45.2 & +16:18:25.0 & K3III       & 4426 & 2.42 & 0.00       \\
HD034255 & 05:20:22.6 & +62:39:13.4 & K4I         & 3000 &      & -0.23      \\
HD126271 & 14:24:18.3 & +08:05:04.6 & K4III       &      &      & -0.12      \\
HD131507 & 14:51:26.4 & +59:17:38.4 & K4III       & 4140 & 1.99 & -0.20      \\
HD149161 & 16:32:36.3 & +11:29:16.9 & K4III       & 3910 & 1.39 & -0.17      \\
HD044274 & 06:21:02.9 & +02:34:07.6 & M           & 4122 & 0.50 & 0.00       \\
HD060501 & 07:35:03.0 & +06:11:39.7 & M           & 4003 & 0.50 & 0.00       \\
HD042543 & 06:12:19.1 & +22:54:30.7 & M0Iab       & 3615 & 0.00 & -0.42      \\
HD039801 & 05:55:10.3 & +07:24:25.4 & M1-M2Ia-Iab & 3547 &      & 0.03       \\
HD035601 & 05:27:10.2 & +29:55:15.8 & M1.5Ia      & 3700 & 0.20 &            \\
HD044537 & 06:24:53.9 & +49:17:16.4 & M2I         & 3055 &      & 0.08       \\
HD078712 & 09:10:38.8 & +30:57:47.3 & M6III       & 3210 & 0.00 & -0.11      \\
HD196610 & 20:37:54.7 & +18:16:06.9 & M6III       &      &      &           \\
\hline
\end{tabular}
\end{table*}

In this work, we have used late-type giant and supergiant stars as templates. The sample stars have been selected from the MUSE and MEGARA stellar libraries presented in \citet{2019A&A...629A.100I} and \citet{2020MNRAS.493..871G} respectively. Tables \ref{tab:MUSE_stars} and \ref{tab:MEGARA_stars} show the characteristics of each of the selected stars listing in columns 1 to 7: (1) the star name; (2,3) the right ascension and declination of the star; (4) its spectral type; (5) its effective temperature; (6) its surface gravity in logarithmic units; and (7) its metallicity.

The Ca II $\lambda \lambda $ 8498, 8542, 8662 \AA\ triplet lines (CaT) have often been used to measure the velocity dispersion of stars in galaxies. The use of these features with respect to other ones present in galaxy spectra presents some advantages. First, the velocity resolution at longer wavelengths is higher than at the bluer part of the spectrum for the same spectral dispersion. At these wavelengths, we expect $\sim $ 37.7 km/s and $\sim$ 6.2 km/s spectral resolutions in $\sigma$ for the MUSE and MEGARA (using the HR-I mode, VPH 863) data respectively. Second, the measurement of these lines is not affected by the presence of TiO bands. And third, the CaT lines are dominated by young supergiant stars, much more luminous than red giants, whose strength increases with decreasing surface gravity \citep{1989MNRAS.239..325D} (although they depend on metal abundance in the low metallicity regime, at moderate to high metallicities surface gravity becomes the dominant parameter governing these strengths). Also, they are less affected by the possible presence of an AGN since the dilution of stellar features due to the presence of a power-law continuum is lower at red wavelengths. Thus, they are especially suitable for the study the velocity dispersions of young star clusters located at the central regions of galaxies, as circumnuclear star-forming regions (CNSFRs).

\begin{figure}
\centering
\includegraphics[width=\columnwidth]{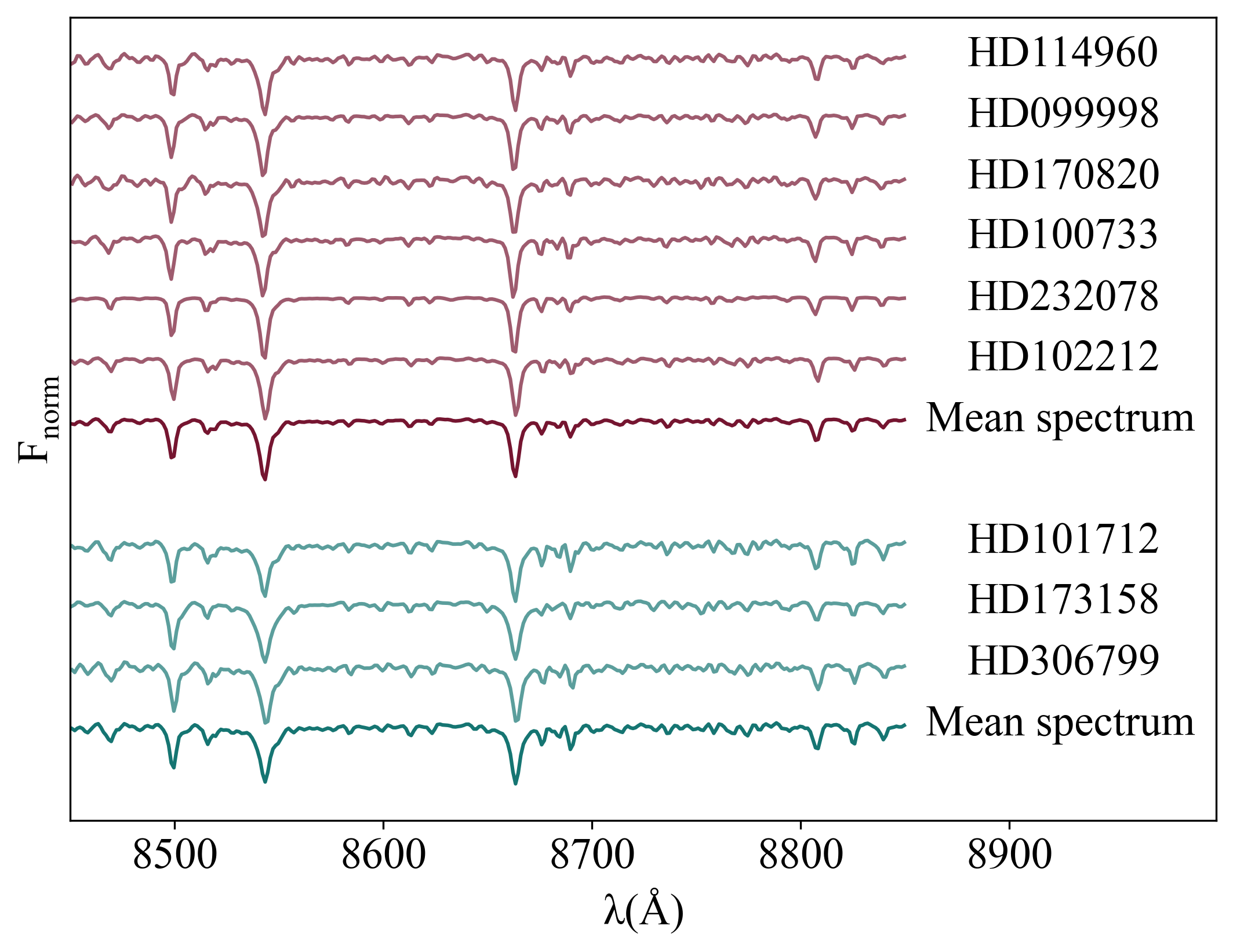}
\caption{MUSE Giant and supergiant stars (upper and lower spectra respectively, see text).}
\label{fig:MUSE_stars}
\end{figure}

\begin{figure}
\centering
\includegraphics[width=\columnwidth]{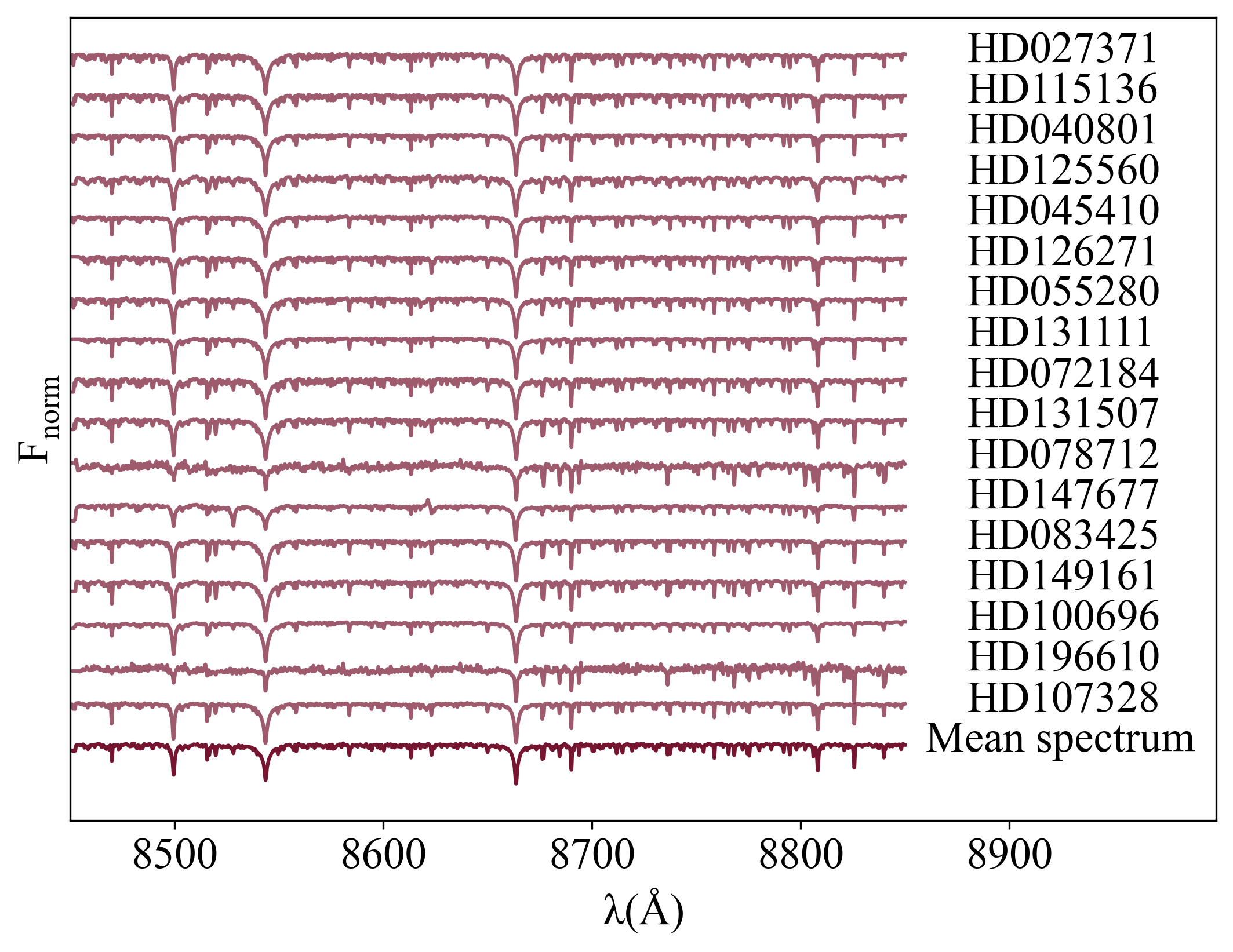}
\includegraphics[width=\columnwidth]{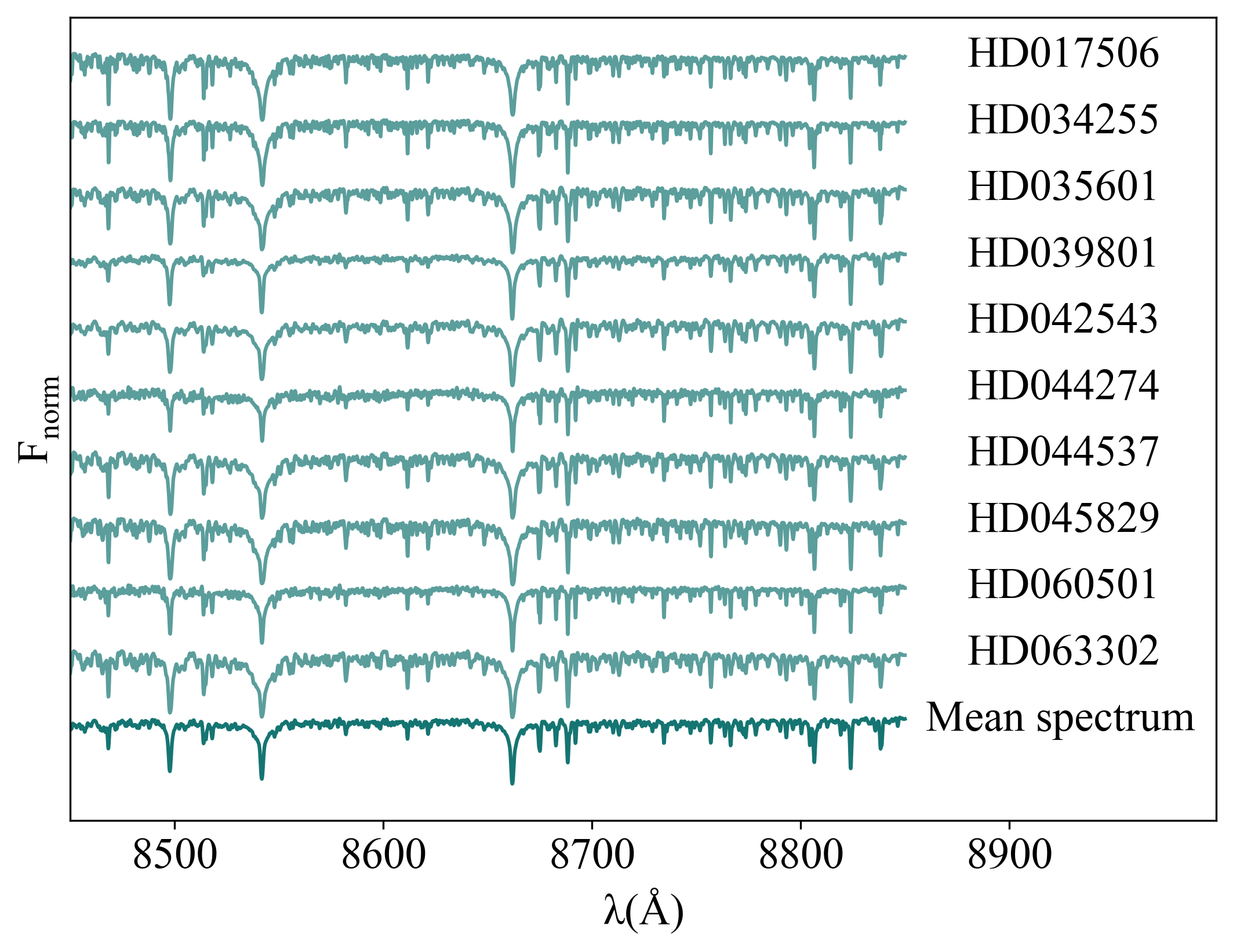}
\caption{MEGARA giant and supergiant stars (upper and lower panels respectively, see text).}
\label{fig:MEGARA_stars}
\end{figure}

Figures \ref{fig:MUSE_stars} and \ref{fig:MEGARA_stars} show the stellar spectral range used in this work. Red and blue spectra correspond to red giant and supergiant stars respectively and at the end of each series the mean spectrum which has been used as template is shown. All the spectra have been corrected for velocity shifts so that no artificial broadening was introduced in the final templates. For our analysis, we have used a mean stellar template computed after aligning all selected spectra and verifying that no apparent broadening is introduced in the procedure. We have checked that the use of different stars of the same spectral type does not introduce errors in the correlation results. Nevertheless, the use of only two stellar types as templates can introduce errors in the velocity dispersion measurements of a given star cluster or galaxy due to the presence of different stellar types with different luminosities. The possible mismatch between the stellar template and the cluster or galaxy lines has been minimised by using the CaT lines since they are very strong in most stars except for the hottest ones. However, it is important to emphasise that our work constitutes only a method revision and its application should be adapted to each particular scientific case.

\section{Cross Correlation Technique}
\subsection{Correlation theoretical concepts} \label{sec:teoria}
The convolution product has been applied with discrete Fourier transforms, using the Convolution, the Correlation and the Parseveral Theorems. For two generic functions, s(t) p(t), these theorems can respectively be abbreviated as: 
\begin{displaymath}
F\left[s(t)\ast p(t)\right](k) = S(k)\cdot P(k)
\end{displaymath}
\begin{displaymath}
F\left[s(t)\otimes p(t)\right](k)=P(k)\cdot S^*(k)
\end{displaymath}
\begin{displaymath}
\displaystyle\sum _{n} s(n)^2 = \frac{1}{N}\displaystyle\sum _{k} |S(k)|^2
\end{displaymath}
\noindent where $F$ denotes the discrete Fourier transform, $\ast$ means the convolution product, $\otimes$ means the cross correlation product, $S(K)$ and $P(k)$ are the Discrete Fourier Transforms of $s(t)$ and $p(t)$ functions respectively, and $S*(K)$ and $|S(k)|$ are the complex conjugate and the module of $S(k)$ respectively. The spectra will be considered periodic in N in order to calculate the discrete Fourier transforms.

The method is based on the idea that galaxy spectra can be represented by a star spectrum convolved with a broadening function: 
\begin{equation}\label{eq:gn}
g(n)\sim \alpha [t(n)\ast b(n-\delta)]
\end{equation}
\noindent where $\ast$ stands for the convolution product, $g(n)$ is the galaxy spectrum, $\alpha $ is the number of stars, $t(n)$ is the template spectrum, $b(n)$ is the broadening function and $\delta$ is the offset of the broadening function with respect to the template. This equation can be understood as the galaxy spectrum being a sum of different star spectra with different velocity offsets.

The broadening function will be assumed to be a Gaussian:
\begin{equation}
b(n) = \frac{1}{\sqrt{2\pi } \sigma} e^{-\frac{n^2}{2\sigma ^2}}
\end{equation}
\noindent whose Fourier transform is: 
\begin{displaymath}
B(k) = \displaystyle\sum _{n} \frac{1}{\sqrt{2\pi } \sigma} e^{-\frac{n^2}{2\sigma ^2}} e^{-\frac{2\pi i n k}{N}} = \frac{e^{-\frac{\left(2 \sigma \pi k\right)^2}{2N^2}}}{\sqrt{2\pi } \sigma}  \displaystyle\sum _{n}  e^{-\frac{\left[n + \left(\frac{2 \sigma^2 \pi i k}{N}\right)\right]^2}{(\sqrt{2}\sigma )^2}}
\end{displaymath}
Taking into account the approximation $\sum e^{n^2/\sigma ^2}\sim \sqrt{\pi} \sigma$ this can be writen as:
\begin{equation}\label{B(k)}
B(k) \sim  e^{-\frac{\left(2 \sigma \pi k\right)^2}{2N^2}}
\end{equation}
Also, a Gaussian amplitude of stellar Fourier transform can be assumed:
\begin{equation}\label{T(k)}
|T(k)| = \sigma _t \left(\frac{2\pi N \tau}{\sqrt{\pi}}\right)^{1/2}e^{-\frac{(2\pi \tau k)^2}{2N^2}}
\end{equation}


\subsection{Cross correlations methods}\label{method}
In this work, we have calculated three cross correlations: (i) the correlation of the galaxy spectrum with itself; (ii) the correlation of the star spectrum with itself; and (iii) the correlation of the galaxy spectrum with the star spectrum. We have used the following normalisation:
\begin{displaymath}
\sigma_g^2 = \frac{1}{N} \displaystyle\sum _{n}g(n)^2 \qquad \sigma_t^2 = \frac{1}{N} \displaystyle\sum _{n}t(n)^2
\end{displaymath}

The correlation of the galaxy spectrum with itself can be calculated as:
\begin{displaymath}
c_{gg}(n) = g(n)\otimes g(n)|_{norm} = \frac{1}{N  \sigma_g^2}\displaystyle\sum _{m} g(m)\cdot g(m-n) 
\end{displaymath}
In the Fourier form:
\begin{equation}\label{eq:c1}
\begin{split}
C_{gg}(k) = F[c_{gg}(n)] = F[g(n)\otimes g(n)]_{norm} 
= \frac{1}{N \sigma_g^2} G(k)\cdot G(k)* \\
=\frac{1}{N \sigma_g^2} |G(k)|^2= \frac{\alpha ^2}{N\sigma_g^2} |T(k)\cdot B(k)|^2
=\frac{\alpha ^2}{N\sigma_g^2} |T(k)|^2\cdot |B(k)|^2\\= \frac{2\pi \tau \alpha^2 \sigma_t^2}{ \sigma_g^2\sqrt{\pi}}\cdot exp\left[{\frac{-k^2}{2 \cdot \left(\sfrac{N}{2\pi\sqrt{2\tau^2 +2\sigma^2}}\right)^2}}\right]
 \end{split}
\end{equation}

The correlation of the star spectrum with itself can be calculated as:
\begin{displaymath}
c_{tt}(n) = t(n)\otimes t(n)|_{norm} = \frac{1}{N  \sigma_t^2}\displaystyle\sum _{m} t(m)\cdot t(m-n) 
\end{displaymath}
In the Fourier form:
\begin{equation}\label{eq:c2}
\begin{split}
C_{tt}(k) = F[c_{tt}(n)] = F[t(n)\otimes t(n)]_{norm} 
= \frac{1}{N \sigma_t^2} T(k)\cdot T(k)*\\
=\frac{1}{N \sigma_g^2} |T(k)|^2=  \frac{2\pi \tau}{ \sqrt{\pi}}\cdot exp\left[\frac{-k^2}{2 \cdot \left(\sfrac{N}{2\pi\sqrt{2}\tau}\right)}\right]
 \end{split}
\end{equation}

Finally, the correlation of the galaxy spectrum with the star spectrum can be calculated as:
\begin{displaymath}
c_{gt}(n) = g(n)\otimes t(n)|_{norm} = \frac{1}{N\sigma_t \sigma_g}\displaystyle\sum _{m} g(m)\cdot t(m-n) 
\end{displaymath}
In the Fourier form:
\begin{equation}\label{eq:c3}
\begin{split}
C_{gt}(k) = F[c_{gt}(n)] = F[g(n)\otimes t(n)]_{norm} 
= \frac{1}{N\sigma_t \sigma_g} G(k) \cdot T^*(k) \\= \frac{1}{N\sigma_t \sigma_g} F[\alpha [t(n)\ast b(n-\delta)]] \cdot T^*(k) 
=\frac{\alpha}{N\sigma_t \sigma_g} B(k)\cdot |T(k)|^2 \\= \frac{2\pi \tau \alpha \sigma_t}{ \sigma_g\sqrt{\pi}}\cdot exp\left[{\frac{-k^2}{2 \cdot \left(\sfrac{N}{2\pi\sqrt{2\tau^2+\sigma ^2}}\right)^2}}\right]
\end{split}
\end{equation}

\subsection{Velocity dispersion from the correlation peak}
A Gaussian behaviour of the correlation peak has been assumed, centered at $\delta$ and with dispersion $\mu$:
\begin{equation}\label{eq:c}
c(n)\sim c(\delta)\cdot e^{-\frac{(n-\delta)^2}{2\mu^2}}
\end{equation}
whose Fourier transform is: 
\begin{displaymath}
C(k) = c(\delta)\displaystyle\sum _{n} 
    e^{-\frac{\delta^2}{2\mu^2}}
    e^{-\frac{\left(n+\frac{2 \mu^2 \pi i k}{N}-\delta \right)^2}{2\mu^2}}
    e^{\frac{\left(\frac{2 \mu^2 \pi i k}{N}-\delta \right)^2}{2\mu^2}}
\end{displaymath}
Taking into account the approximation $\sum e^{n^2/\sigma ^2}\sim \sqrt{\pi} \sigma$ it can be writen as:
\begin{equation}\label{eq:C(k)}
C(k) \sim  c(\delta) \sqrt{2\pi} \mu_i \cdot e^{-\frac{2  \pi i k \delta_i}{N}} \cdot exp\left[-\frac{k^2}{2\left(\sfrac{N}{2\pi \mu } \right)^2}\right] 
\end{equation}

The comparison between this last equation and the results of the three calculated cross-correlation functions (see Sec. \ref{method}) establishes relationships between the correlation peaks, $\mu _{gg,tt,gt}$, the template width in the Fourier space, $\tau$, and the broadening function width,  $\sigma$. From Eqs. \ref{eq:c1} and \ref{eq:C(k)}:
\begin{equation}\label{eq:mu1}
\mu _{gg}^2 = 2\tau^2+2\sigma^2
\end{equation}
From Eqs. \ref{eq:c2} and \ref{eq:C(k)}:
\begin{equation}\label{eq:mu2}
\mu _{tt}^2 = 2\tau^2
\end{equation}
From Eqs. \ref{eq:c3} and \ref{eq:C(k)}:
\begin{equation}\label{eq:mu3}
\mu _{gt}^2 = 2\tau^2+\sigma^2
\end{equation}

Combining these three last equations, we can fit a Gaussian to the correlation functions and derive the  broadening function width, which is related to the velocity dispersion to be measured. From Eqs. \ref{eq:mu1} and \ref{eq:mu2}:
\begin{equation}\label{eq:sigma1}
\sigma^2 = \frac{\mu^2_{gg}-\mu^2_{tt}}{2}
\end{equation}
From Eqs. \ref{eq:mu1} and \ref{eq:mu3}:
\begin{equation}\label{eq:sigma2}
\sigma^2 = \mu^2_{gg}-\mu^2_{gt}
\end{equation}
From Eqs. \ref{eq:mu2} and \ref{eq:mu3}:
\begin{equation}\label{eq:sigma3}
\sigma^2 = \mu^2_{gt}-\mu^{2}_{tt}
\end{equation}
This last equation is the one proposed by \citet{1979AJ.....84.1511T}. The other two equations are equivalent to this one being their assumptions and mathematical development analogous to the original method.

\subsection{Spectrum preparation for the correlation analysis}
Each spectrum analysed with the cross-correlation technique should be binned into logarithmic wavelengths in order to get a uniform velocity shift. Let $\lambda$ be the wavelength of the spectrum and n the equivalent bin number; then we can apply the following equation: 
\begin{equation}\label{eq:1}
n(\lambda) = A\cdot ln(\lambda)+B
\end{equation}
where A and B are constants for each spectrum. 

The number of the sampled bins depends on the velocity resolution, $\Delta v$, which we need to obtain. If the wavelength range of the analysed spectrum is between $\lambda _1$ and $\lambda _2$, the minimum number of bins that should be used to not lose spectral resolution can be calculated as: 
\begin{equation}\label{eq:N}
N = \frac{ln(\lambda_ 2/\lambda_ 1)}{ln(\Delta v/c +1)}
\end{equation}
where c is the speed of light in the same units of $\Delta v$. Using this value, we can calculate the A and B constant values from Eq. \ref{eq:1}:
\begin{equation}\label{eq:B}
B = N\cdot \left(1-\frac{ln(\lambda_ 2)}{ln(\lambda_ 1)}\right)
\end{equation}
\begin{equation}\label{eq:A}
A = -\frac{N}{ln(\lambda _1)}\cdot \left(1-\frac{ln(\lambda_ 2)}{ln(\lambda_ 1)}\right)
\end{equation}

Any emission line present in the spectrum should be removed. Also, the continuum spectrum should be subtracted. A second order polynomial is sufficient to model this component since any mistake made will be corrected in the following steps.

The final step in the spectrum preparation is filtering the high and low frequency variations that can affect the velocity dispersion determination. They are associated with noise components and errors in the continuum subtraction respectively. This filtering is performed by applying a band-pass to the Fourier spectrum transform with  minimum and maximum wave numbers, k$_{min}$ and k$_{max}$ which can be calculated as follows:
\begin{equation}\label{eq:kmin}
k_{min} = \frac{N}{2\pi A \cdot ln\left(1-\Delta \lambda^{max}/\lambda _1\right) }
\end{equation}
\begin{equation}\label{eq:kmax}
k_{max} = \frac{N}{2\pi A \cdot ln\left(1-\Delta \lambda^{min}/\left(2 \cdot \lambda _2 \sqrt{2\cdot ln(2)}\right) \right)}
\end{equation}
\noindent where $\Delta \lambda ^{min}$ and $\Delta \lambda ^{max}$ are the minimum and maximum detail that we expect in each spectrum. We have assumed as the minimum value the nominal spectral resolution in \AA , and as the maximum value 10 \AA, a size significantly larger than the width of the absorption lines in each spectrum.

\section{Analysis and results}

We have used the mean spectrum of all stars presented above to calculate the broadening function of MUSE and MEGARA data aiming to check: (i) the traditional methodology proposed by Tonry and Davis \citep[][from Eq. \ref{eq:sigma3}]{1979AJ.....84.1511T} and (ii) the other two proposed in this work (from Eq. \ref{eq:sigma1} and Eq. \ref{eq:sigma2}). Additionally, we have analysed the results of the frequency filtering in the data and we have quantified the effects of the spectrum preparation steps in the results.

\subsection{Application to MUSE data}

First, we have selected wavelengths between 8450\AA\ and 8850 \AA\ in order to exclude the [OI]$\lambda $ 8446 \AA\ and Pa11 lines. We have binned the template spectrum into logarithmic wavelengths using N = 512 bins which corresponds to a uniform velocity shift of $\Delta $v = 27.1 km/s, lower than the nominal velocity resolution of MUSE. Using Eqs. \ref{eq:A} and \ref{eq:B} we have evaluated the constants as A = 11070.03 and B = -100094.31.

Next, we have calculated and subtracted the continuum spectrum, also masking emission nebular lines and absorption stellar lines using widths of 6 \AA\ and 20 \AA\ respectively, and we have fitted a quadratic equation. 

Later, we have used the N value and the selected wavelengths to calculate the band-pass filter constants using Eqs. \ref{eq:kmin} and \ref{eq:kmax}. The selected wave numbers in the Fourier space for MUSE data are $k_{min}$ $\sim$ 3 and $k_{max}$ $\sim$ 60 which correspond to a bin number shift between 1.4 and 26.2. Thus, the Fourier transform of each spectrum  has been multiplied by the following function: 
\begin{equation}
    f(k) =
\begin{cases*}
    0 & if k < 1.5\\\
    k/1.5-1 & if 1.5< k < 3\\\
    1 & if 3< k < 60\\\
    -k/60+2 & if 60< k < 120\\\
    0 & if k > 120
\end{cases*}
\end{equation}

\begin{figure}
\includegraphics[width=\columnwidth]{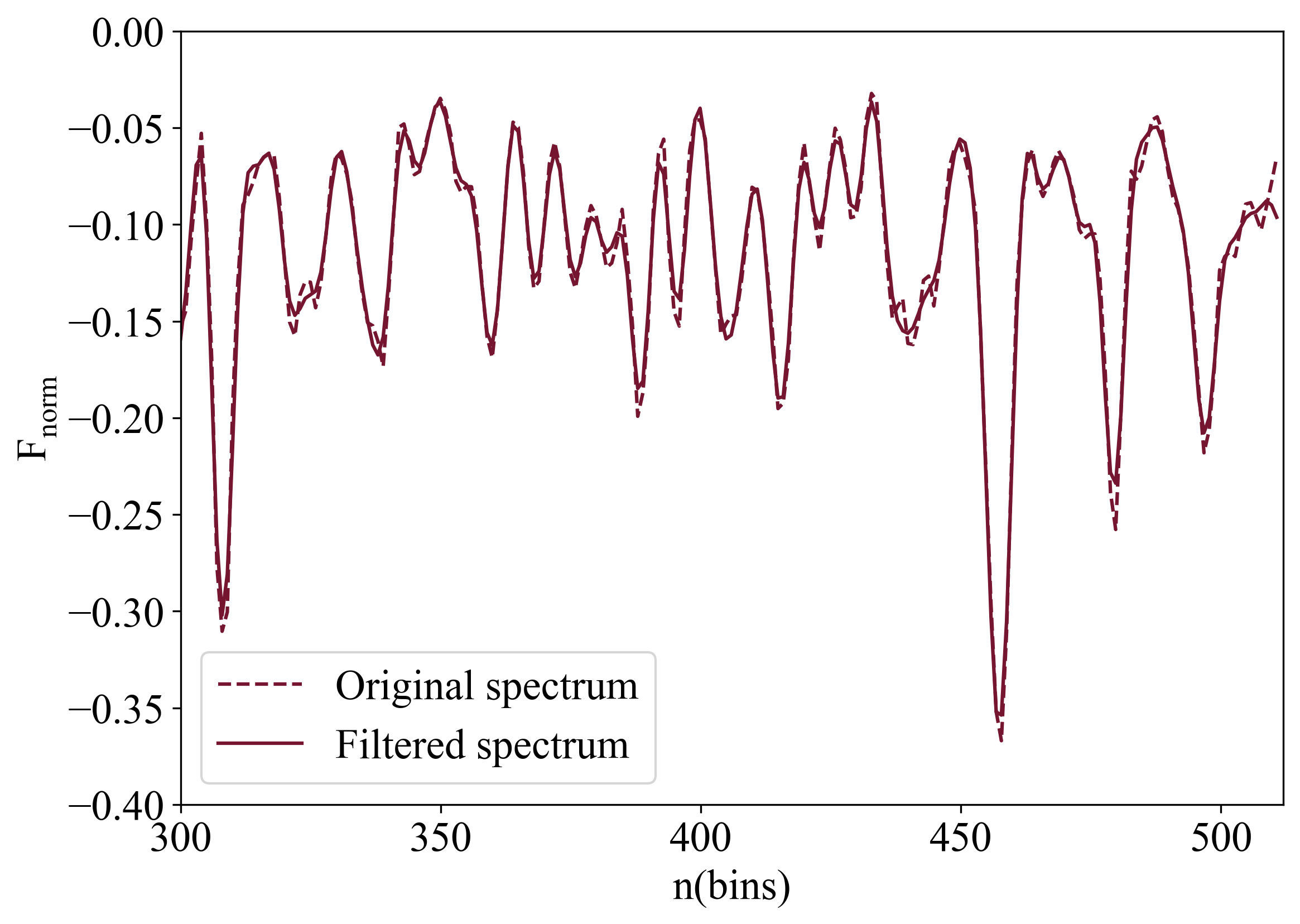}
\caption{MUSE star spectrum after resampling into logarithmic wavelengths and applying the pass-band filter for the higher frequencies.}
\label{fig:filter_frec_MUSE}
\end{figure}

Figure \ref{fig:filter_frec_MUSE} shows the result of this filtering at high frequencies. The original spectrum and the filtered one are shown as dotted and solid lines respectively. We can see that the filter application removes the noise component of the analysed spectrum.

\begin{figure}
\includegraphics[width=\columnwidth]{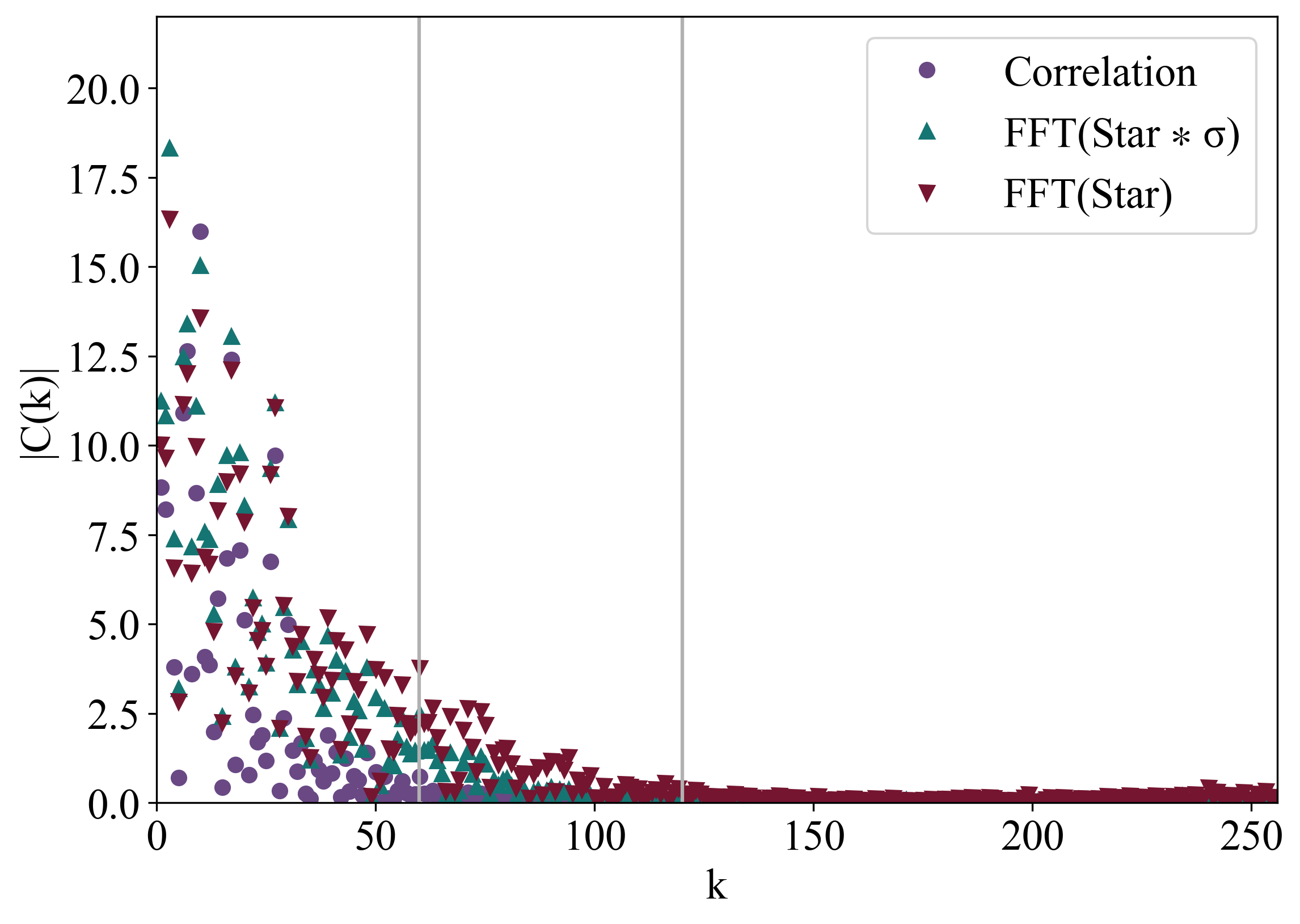}
\includegraphics[width=\columnwidth]{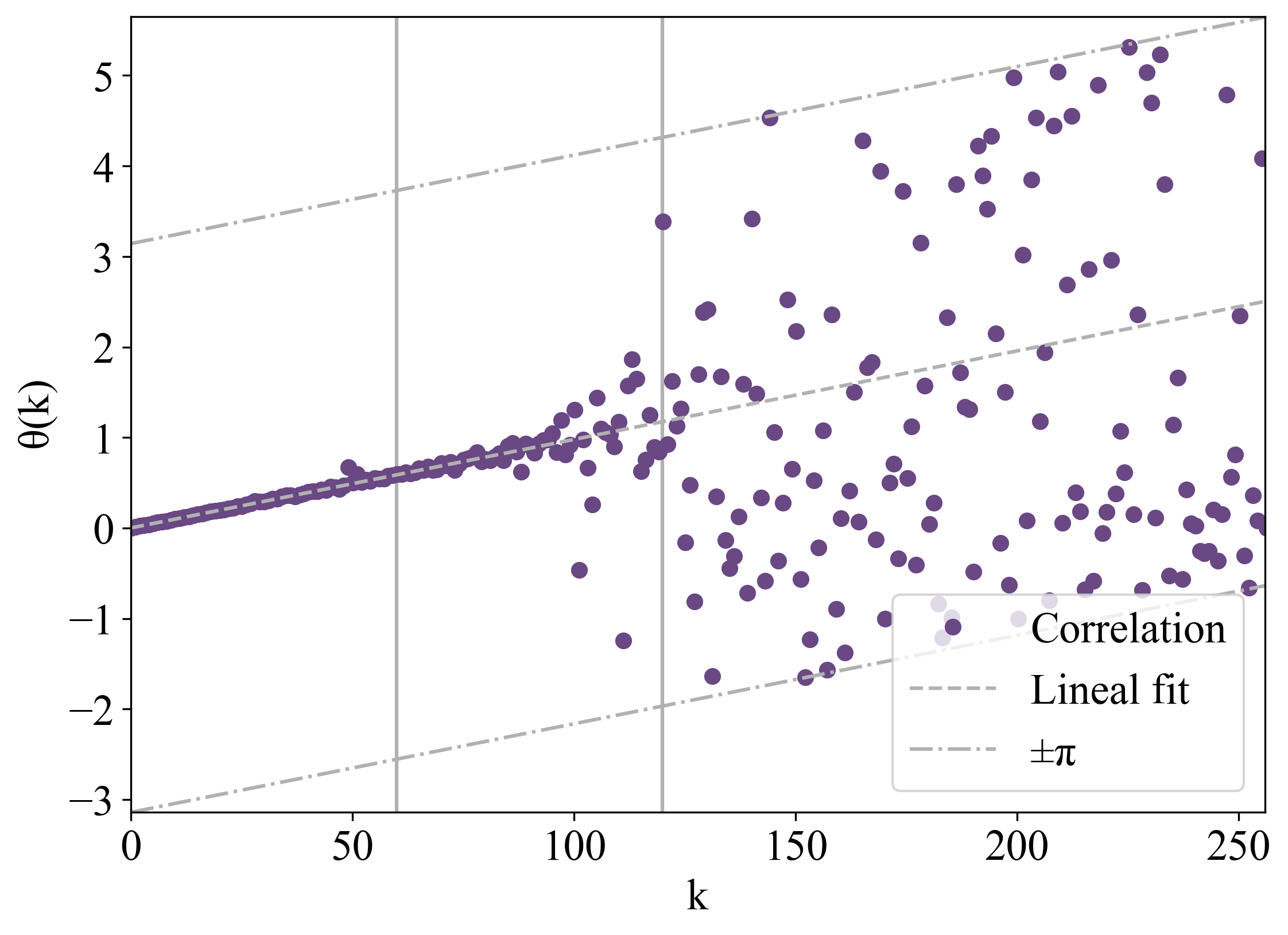}
\caption{Upper panel: Amplitude of the Fourier transform of the MUSE stellar template (red triangles), this template convolved with a Gaussian function (blue triangles) and the cross correlation of the two (purple dots). Lower panel: Fourier transform phase of the same functions with the same colour coding.  }
\label{fig:amplitude_phase_MUSE}
\end{figure}

In order to verify if the frequency filter is well applied, we have simulated a cluster spectrum convolving the stellar template with a known Gaussian function of width $\sigma$ = 1.09 \AA, which corresponds to the MUSE spectral resolution. Figure \ref{fig:amplitude_phase_MUSE} shows, in the upper and lower panels respectively, the phase and the amplitude of the cross-correlation function in the Fourier space as a function of wave number. Red triangles show the template results; blue triangles show the results of the simulated cluster spectrum; and purple dots show the results of the correlation between the two. Vertical grey lines mark the  k$_{max}$ and 2$\cdot$k$_{max}$ values used for the band-pass filter. We can see that the phase becomes random at higher values of the 2$\cdot$k$_{max}$ value while the amplitude is compatible with zero in the same range. Hence, we can conclude that the chosen k$_{max}$ values are correct.

\subsection{Application to MEGARA data}
\label{sec:megara_apli}

In this case, we have selected the same wavelength range although the needed number of bins is now larger, N = 4096 ($\Delta $v = 3.4 km/s), corresponding to a value lower than the velocity resolution of HR-I MEGARA setup. The appropriate binning constants are A = 88560.21 and B = -800754.52. As in the case of MUSE data, nebular and stellar lines have been removed.

Then, we have applied the band-pass filter with wave numbers in Fourier space $k_{min}$ $\sim$ 3 and $k_{max}$ $\sim$ 350 corresponding to bin number shifts between 1.9 and 209.4: 
\begin{equation}
    f(k) =
\begin{cases*}
    0 & if k < 1.5\\\
    k/1.5-1 & if 1.5< k < 3\\\
    1 & if 3< k < 350\\\
    -k/350+2 & if 350< k < 700\\\
    0 & if k > 700
\end{cases*}
\end{equation}

\begin{figure}
\includegraphics[width=\columnwidth]{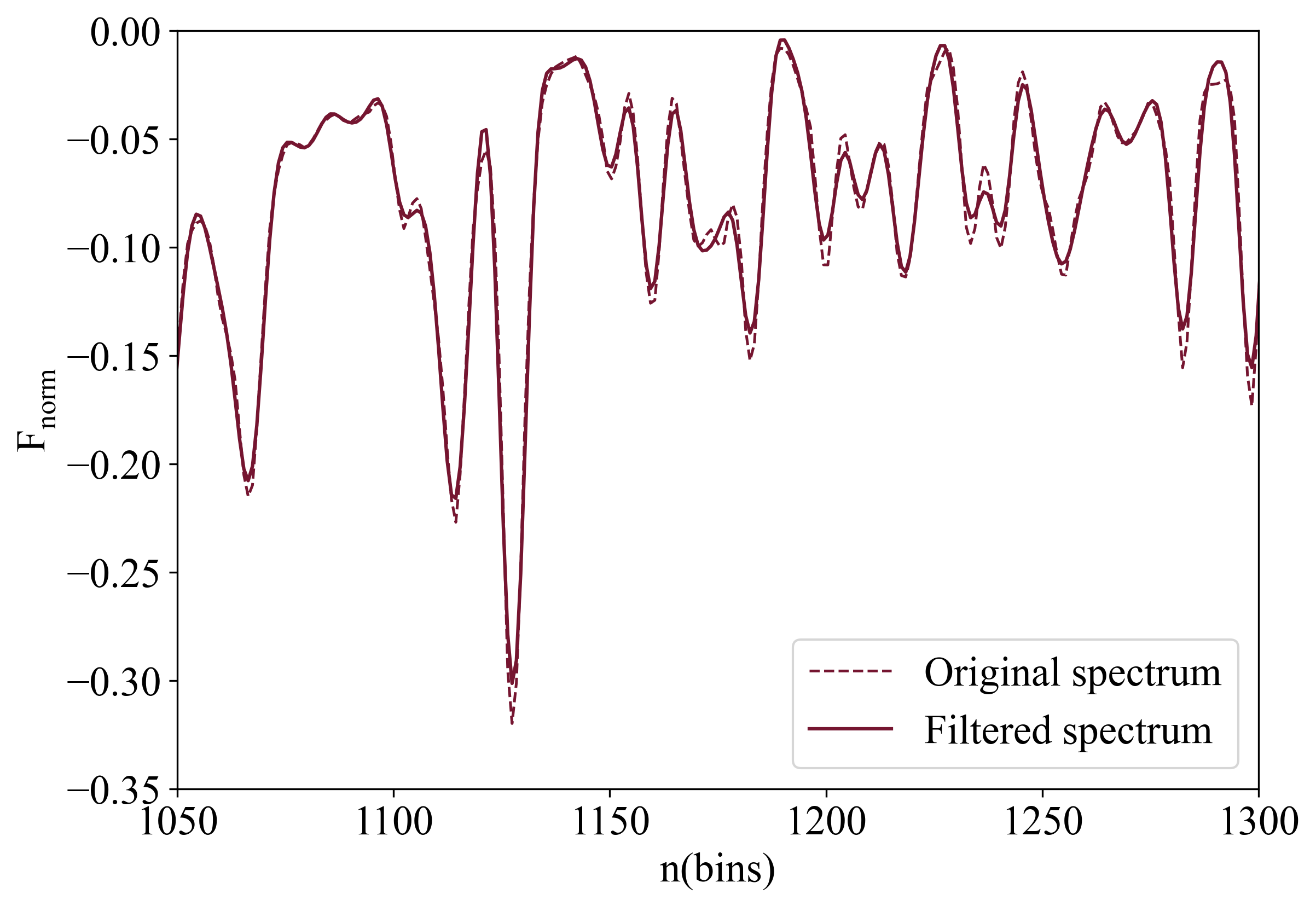}
\includegraphics[width=\columnwidth]{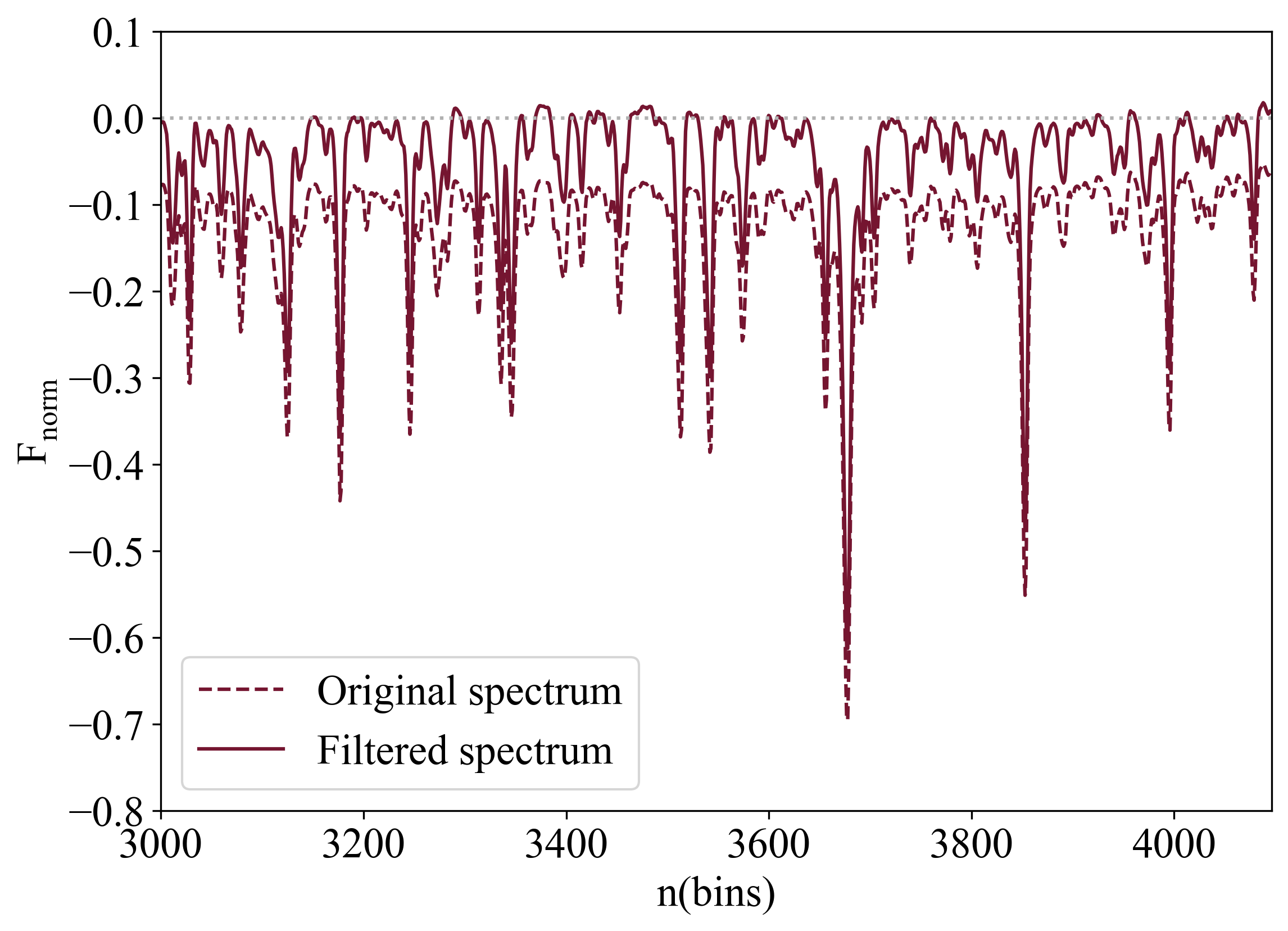}
\caption{MEGARA star spectrum after resampling into logarithmic wavelengths and applying the pass-band filter for the higher frequencies (upper panel) and the lower ones (bottom panel).}
\label{fig:filter_frec_MEGARA}
\end{figure}

Figure \ref{fig:filter_frec_MEGARA} shows the result of this filtering on the data. The upper panel shows the effect of the high frequency variation whose behaviour is similar to the case of MUSE data. The lower panel shows the effect of the low frequency filter which corrects the errors in the continuum subtraction as expected.

\begin{figure}
\includegraphics[width=\columnwidth]{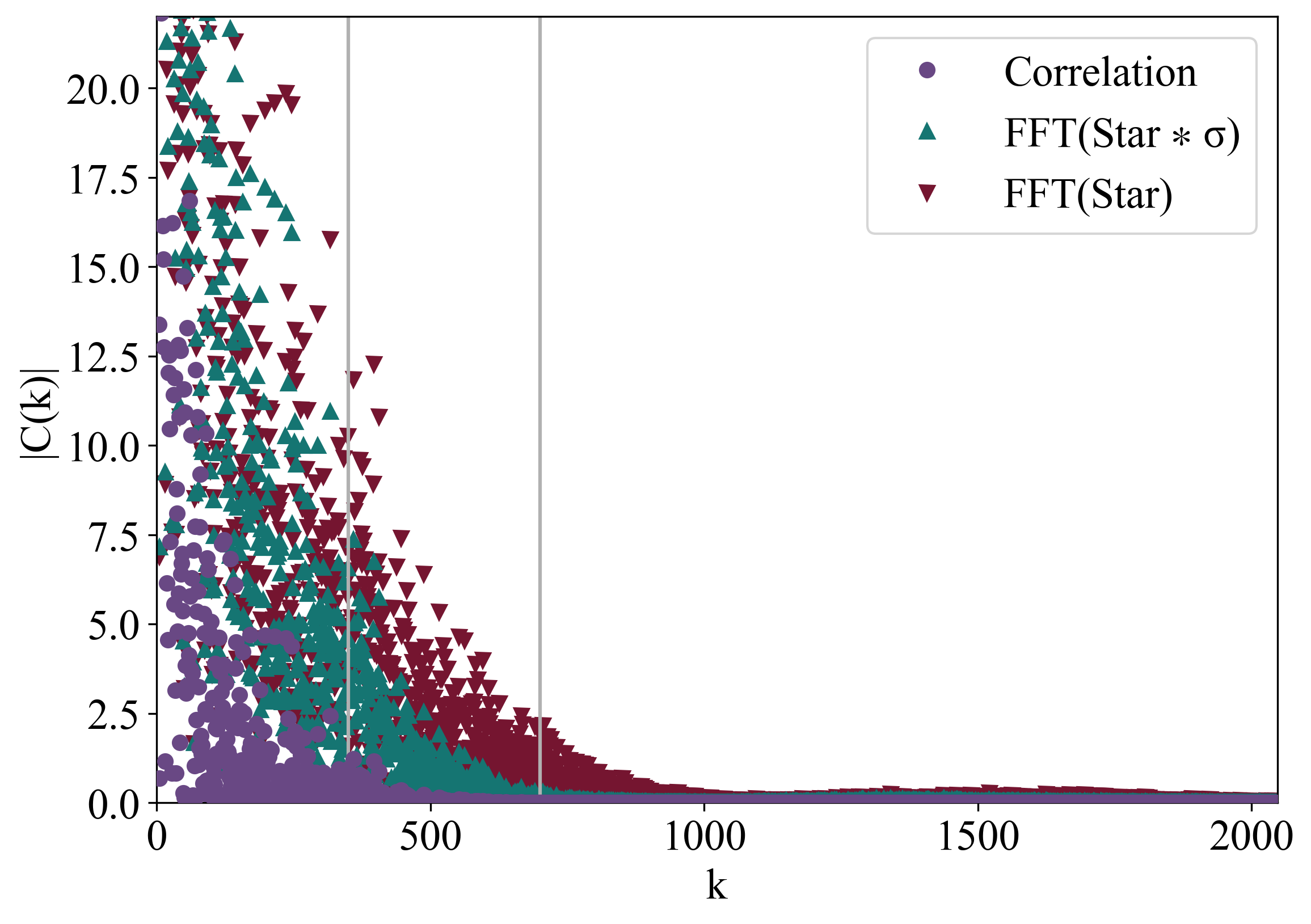}
\includegraphics[width=\columnwidth]{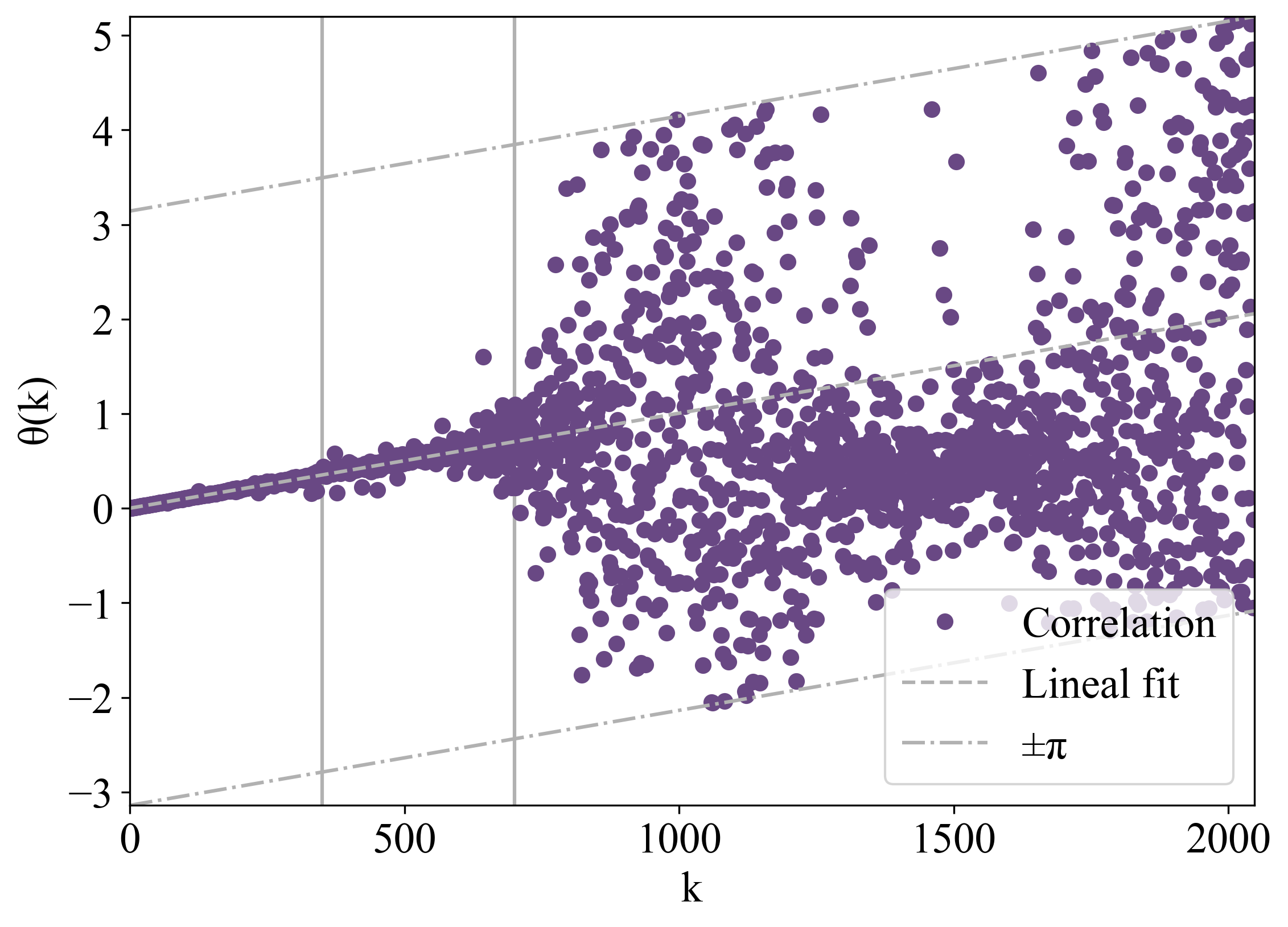}
\caption{Upper panel: Amplitude of the Fourier transform of the MEGARA stellar template (red triangles), this template convolves with a Gaussian function (blue triangles) and the cross correlation of the two previous ones (purple dots). Lower panel: Phase of the same functions mentioned above. The color code is the same as in the upper panel.}
\label{fig:amplitude_phase_MEGARA}
\end{figure}

In this case, we have convolved the stellar template with a Gaussian function with $\sigma$ = 0.18 (MEGARA spectral resolution, HR-I setup) to check if the filter application is correct. Figure \ref{fig:amplitude_phase_MEGARA} shows the phase (upper panel) and the amplitude (lower panel) of the cross-correlation function (the color code is the same as in Fig. \ref{fig:amplitude_phase_MUSE}). Vertical grey lines mark the values used in the band-pass filter (k$_{max}$ = 350 and 2$\cdot$k$_{max}$ = 700 values). This result is similar to that found for MUSE data: at higher wave number values the phase is random and the amplitude of the correlation function is compatible with zero.

\section{Discussion}

\subsection{Method assumptions at different spectral resolutions}\label{sec:suposiciones}
The main purpose of this work is to test the method and assumptions of the classical cross-correlation technique. The principal assumptions of this method are: (i) the broadening function is assumed to be a Gaussian; (ii) the cross-correlation function has a Gaussian behavior and (iii) the amplitude of stellar Fourier transform can also be represented by a Gaussian. 
The first one depends on the particular scientific case to which this methodology is applied since it is based on the idea that the analysed spectrum can be well represented by a star spectrum convolved with a broadening function. This is true, for example, in the case of Seyfert galaxies, circumnuclear star-forming regions and HII regions \citep[][]{2006ApJ...641..117G,2007MNRAS.378..163H,1995ApJS...99...67N,2001BSAO...51...11M}.

\begin{figure}
\includegraphics[width=\columnwidth]{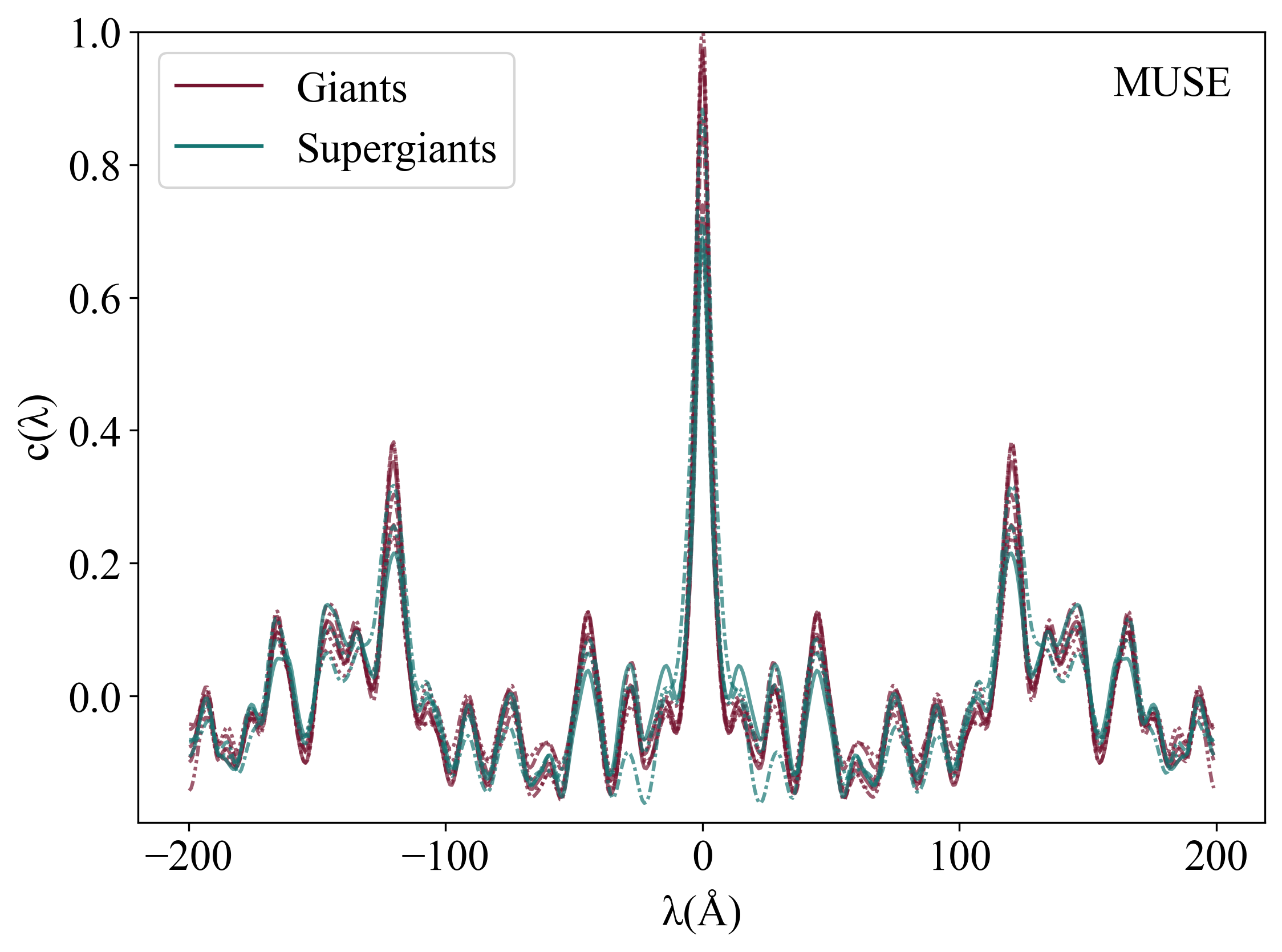}
\includegraphics[width=\columnwidth]{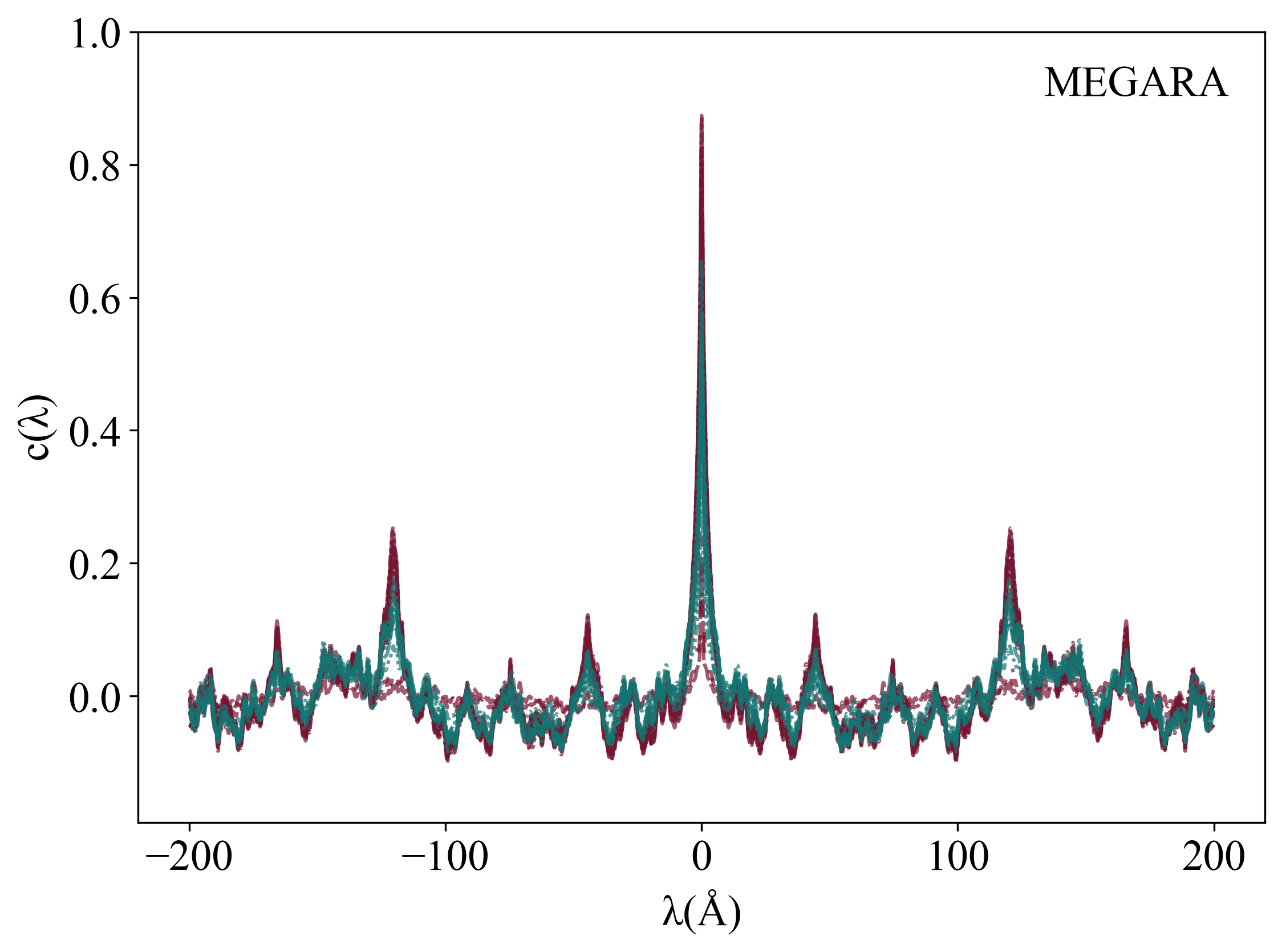}
\caption{Autocorrelation functions for each star from MUSE (upper panel) and MEGARA (lower panel) data. Giant and supergiant stars are marked with red and blue lines respectively.}
\label{fig:pruebas_corr_stars}
\end{figure}

In order to evaluate the second assumption, we have correlated each stellar template with itself;  this cross correlation is the one carrying the instrumental resolution information and therefore is usually narrower than the star-to-galaxy or galaxy-to-galaxy cross-correlation, hence in very high resolution data it can become too narrow to be adequately represented by a Gaussian function.  

Figure \ref{fig:pruebas_corr_stars} shows the results obtained for all the stars used in this work observed with the MUSE and MEGARA instruments (upper and lower panels respectively). First, we can see that all the stars observed with each of the instruments present similar widths and therefore we are not introducing errors by using their mean value as the star template in our analysis. Then, it is also important to emphasise that MUSE star-to-star cross correlation functions present a Gaussian behaviour and hence the assumption (ii) of the method is correct and therefore applicable at this instrumental resolution. However, the results obtained for MEGARA data show just the opposite: the correlation peak is too narrow and  cannot be approximated by a Gaussian function. Thus, the application of the Tonry \& Davis' method to this kind of data would overestimate the obtained velocity dispersion and any quantity derived from it, including the calculation of dynamical masses.

\begin{figure}
\includegraphics[width=\columnwidth]{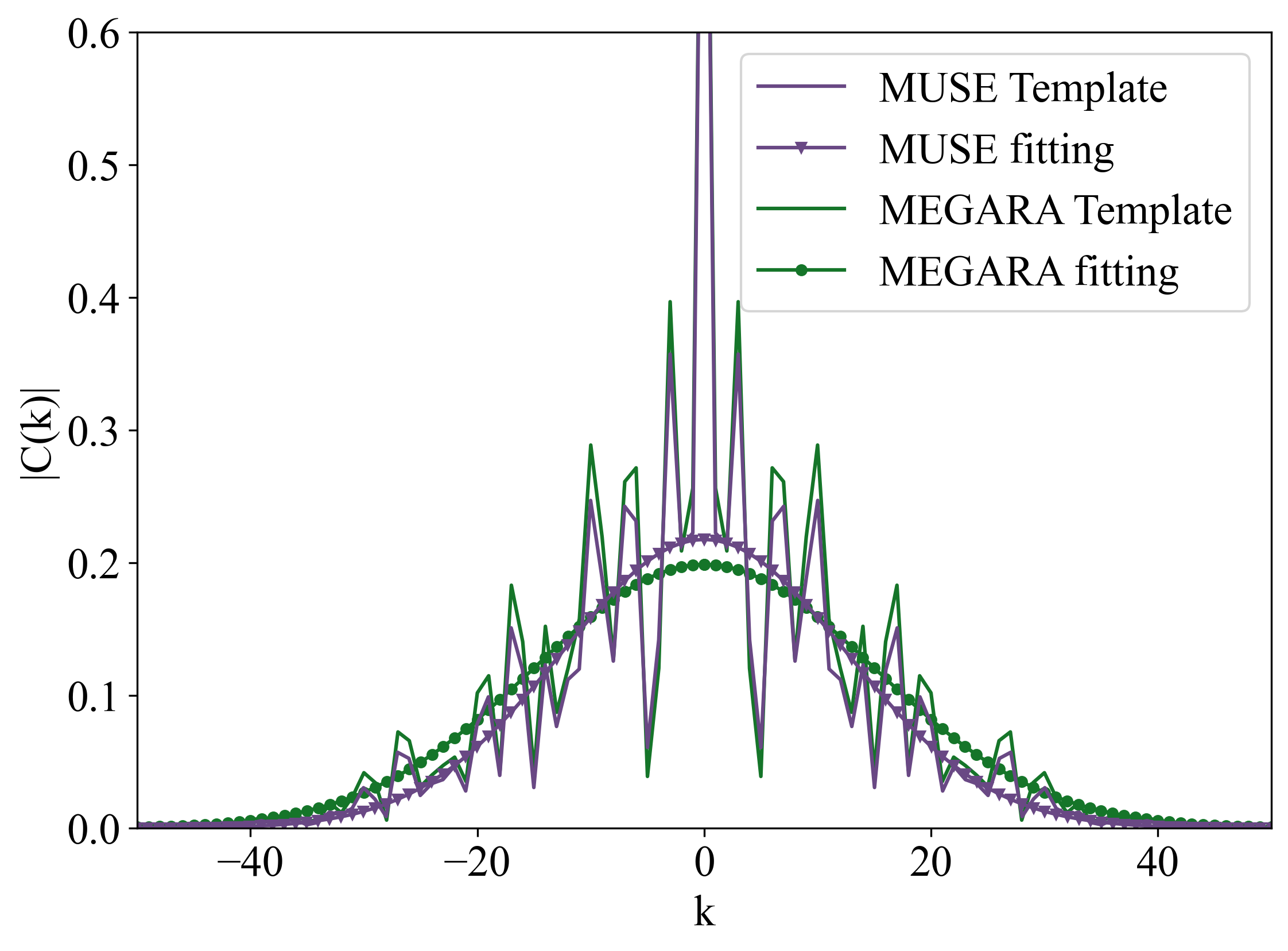}
\caption{Amplitude of the Fourier transform of the MUSE (purple line) and MEGARA (green line) stellar templates. Lines made of triangles and dots show the Gaussian fits performed for these two functions respectively.}
\label{fig:pruebas_amp_stars}
\end{figure}

Finally, Figure \ref{fig:pruebas_amp_stars} shows the Fourier transform amplitudes of MUSE and MEGARA templates (in purple and green respectively); triangle and dot lines correspond to the two fitted Gaussian functions. The central peak of each one corresponds to the low frequency variations which are removed from the correlation function after the band-pass filtering. We can see that in both cases a Gaussian amplitude can be assumed, thus the assumption (iii) remains valid at high spectral resolutions.

\subsection{Comparison among methods}

We have verified the spectral resolution range at which the traditional correlation method and the two ones proposed in this work are valid. In order to do that, we have convolved each stellar template spectrum  with known Gaussian functions of different $\sigma$ values, simulating velocity dispersions from 10 km/s to 200 km/s in steps of $\sim$ 10 km/s for MUSE data and from 2 km/s to 100 km/s in steps of $\sim$ 5 km/s for MEGARA data. These spectra have been then cross-correlated with the original template and the output broadening function widths have been calculated. 

\begin{figure}
\includegraphics[width=\columnwidth]{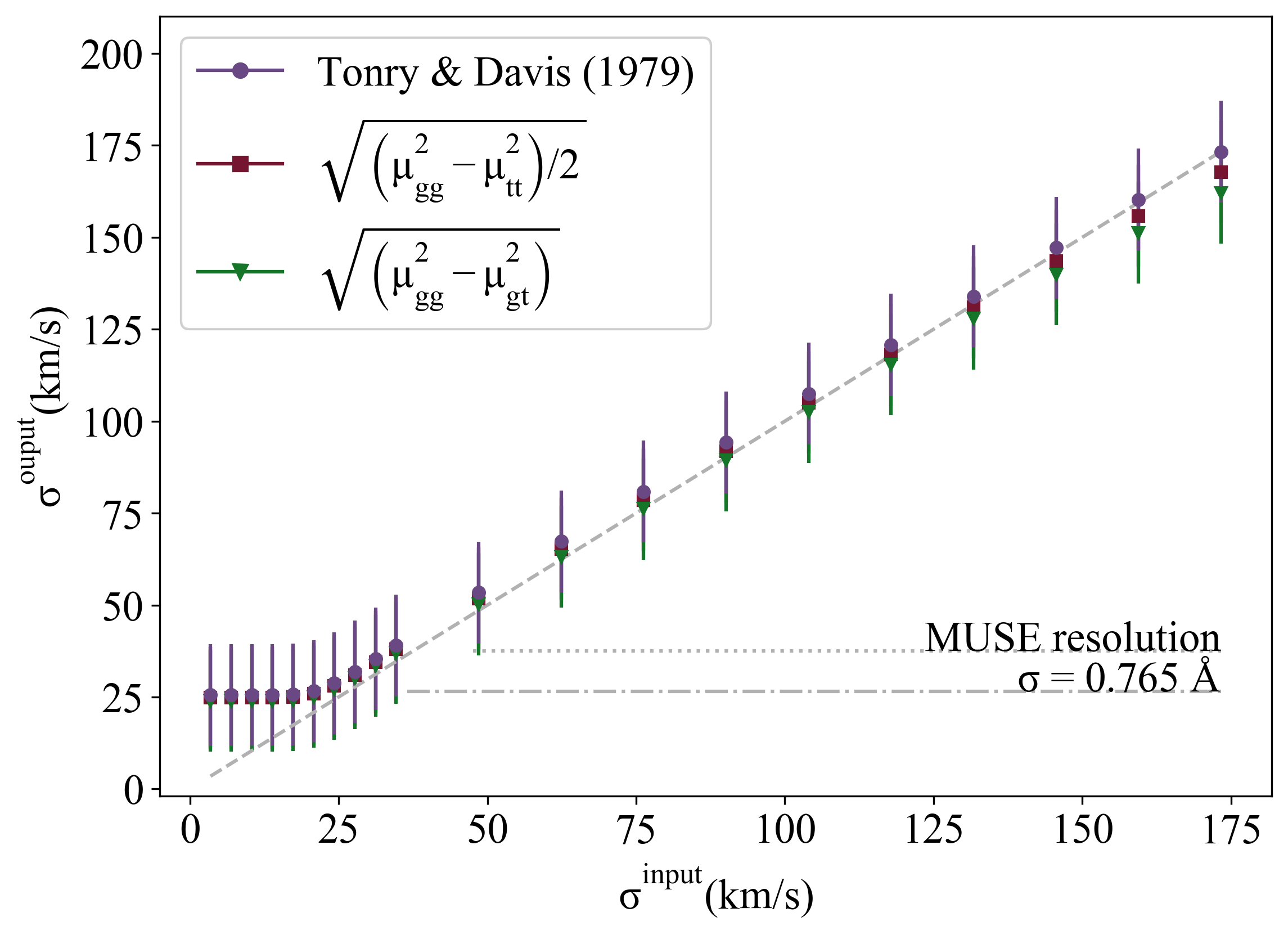}
\includegraphics[width=\columnwidth]{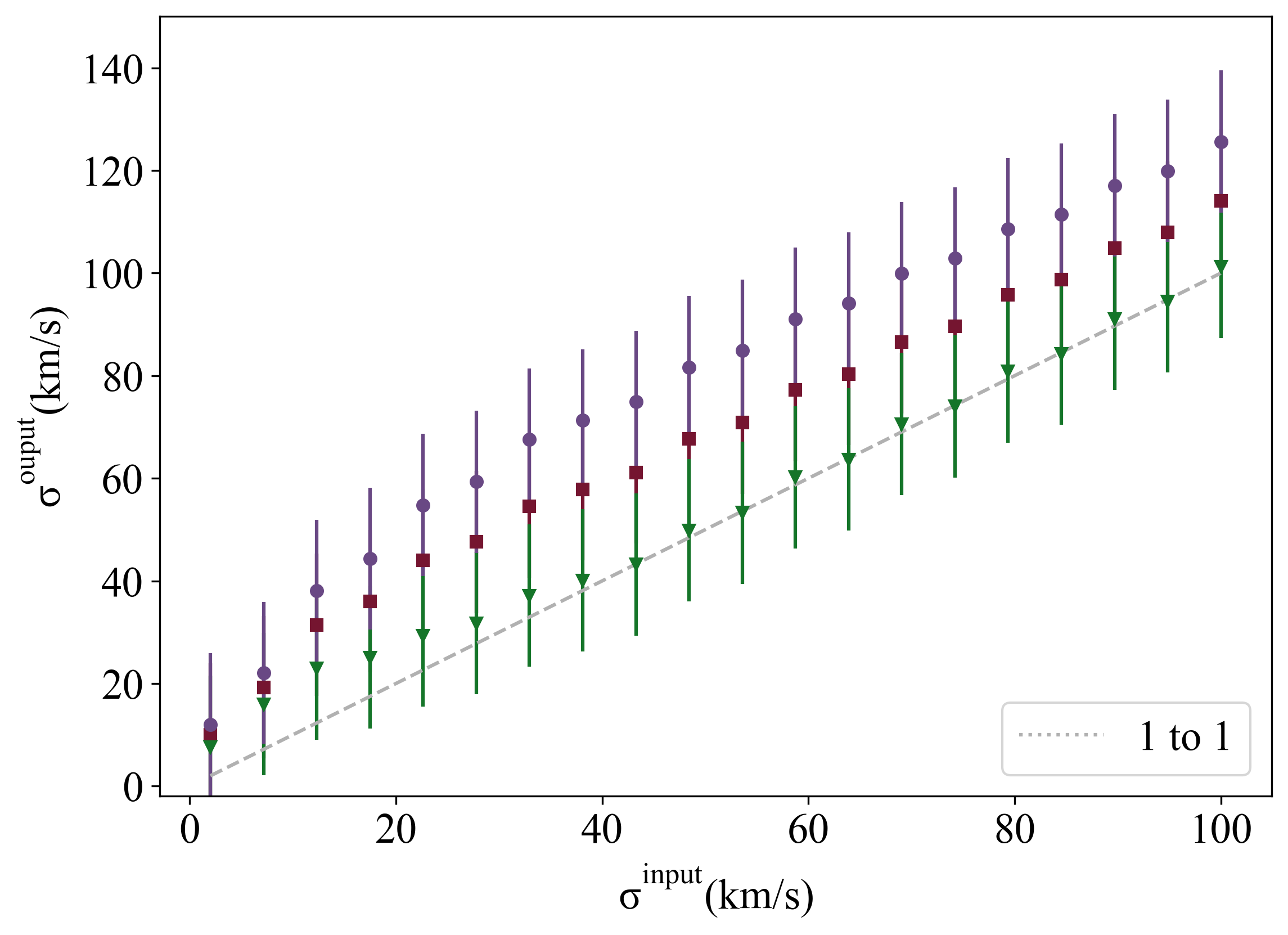}
\caption{Velocity dispersion obtained from the three cross correlation methods studied in this work: the traditional Tonry \& Davis' method (purple dots), the one obtained from [g $\otimes$ g] and [t $\otimes$ t] (red squares) and from [g $\otimes$ g] and [g $\otimes$ t] (green triangles). The results from MUSE and MEGARA data are shown in the upper and lower panel respectively. Error bars have the same length for all the points. }
\label{fig:pruebas_gaussianas}
\end{figure}

The upper panel of Figure \ref{fig:pruebas_gaussianas} shows the comparison between input and output broadening function widths for the three methods proposed in Sec. \ref{sec:teoria} for MUSE data. Purple dots show Tonry \& Davis' method results; red squares show the results using the galaxy-to-galaxy and star-to-star cross correlations, and green triangles show the results in the case of using the galaxy-to-galaxy and galaxy-to-star cross correlations. We can see, at the MUSE spectral resolution the results are entirely compatible within the errors for the tree methods which confirms the results of the previous section that Tonry \& Davis' method can be safely applied at MUSE spectral resolution. Also an asymptotic behavior is found at $\sigma$ = 0.765 \AA\ ($\Delta$v = 26.51 km/s) that probably corresponds to the empirical spectral resolution of the data, that is somewhat lower than the nominal one ($\sigma$ = 2.556 \AA, $\Delta$v = 37.62 km/s). 

The lower panel of the figure shows the same comparison described above but for MEGARA data. In this case, we can see that the two methods which include the star-to-star correlation are not able to recover the input velocity dispersion. At this spectral resolution Tonry \& Davis' method fails being the most discrepant one. In fact it overestimates the derived velocity dispersions by factors ranging from 25 \% at a sigma of 100 km/s to 220 \% at a sigma of 12 km/s. These discrepancies would propagate to any derived quantity. In particular, dynamical masses scale with the square of the velocity dispersion, hence for velocity dispersions of 100 km/s the derived masses would be overestimated by about 50\%; however at low velocity dispersions, as the ones being sought with the use of high resolution data, the derived masses would be overestimated by about one order of magnitude. The best technique to use is the one including the two broader correlation functions, g(n) $\otimes$ g(n) and g(n) $\otimes$ t(n), and, even in this case, discrepancies begin to appear at the lower velocity dispersion values, although they are always within the errors. Therefore this is the recommended method to be applied for high spectral resolution data and results found by the usual application of Tonry \& Davis' method cannot be considered as correct.

\begin{figure}
\includegraphics[width=\columnwidth]{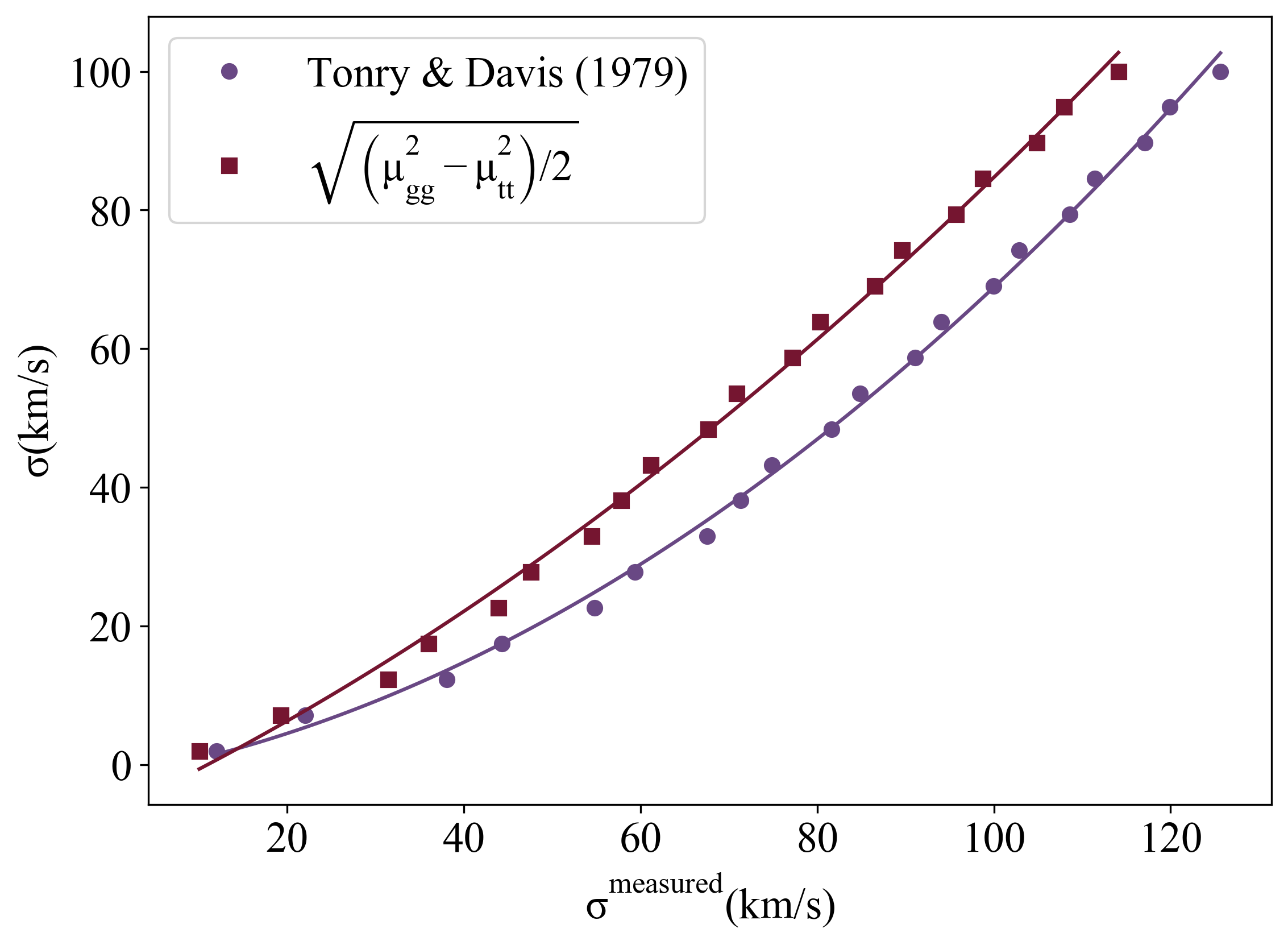}
\caption{Velocity dispersion correction for MEGARA data for the three cross correlation methods studied in this work: the traditional Tonry \& Davis (purple dots) and the ones obtained from [g $\otimes$ g] and [t $\otimes$ t] (red squares) and  [g $\otimes$ g] and [g $\otimes$ t] (green triangles).}
\label{fig:corection}
\end{figure}

We have used Fig. \ref{fig:pruebas_gaussianas} to quantify the differences between measured and real velocity dispersions in the two methods which include the star-to-star correlation. Fig. \ref{fig:corection} shows these differences. A second order polynomial has been fitted for each method, from 10 km/s to 120 km/s, in order to estimate a correction for the MEGARA spectral resolution. For the traditional Tonry \& Davis' method, the correction applied can be calculated using the following equation:
\begin{equation}
\sigma = -1.8339+0.2212\cdot \sigma_{measured} + 0.004858 \cdot (\sigma_{measured})^2
\end{equation}
where $\sigma$ is the real velocity dispersion in km/s and $\sigma_{measured}$ is the measured one in the same units. For the method [g $\otimes$ g] - [t $\otimes$ t] the correction is given by:
\begin{equation}
\sigma = -6.9671+0.6014\cdot \sigma_{measured} + 0.003154 \cdot (\sigma_{measured})^2
\end{equation}

\subsection{Error analysis}

\begin{figure*}
\centering
\includegraphics[width=\textwidth]{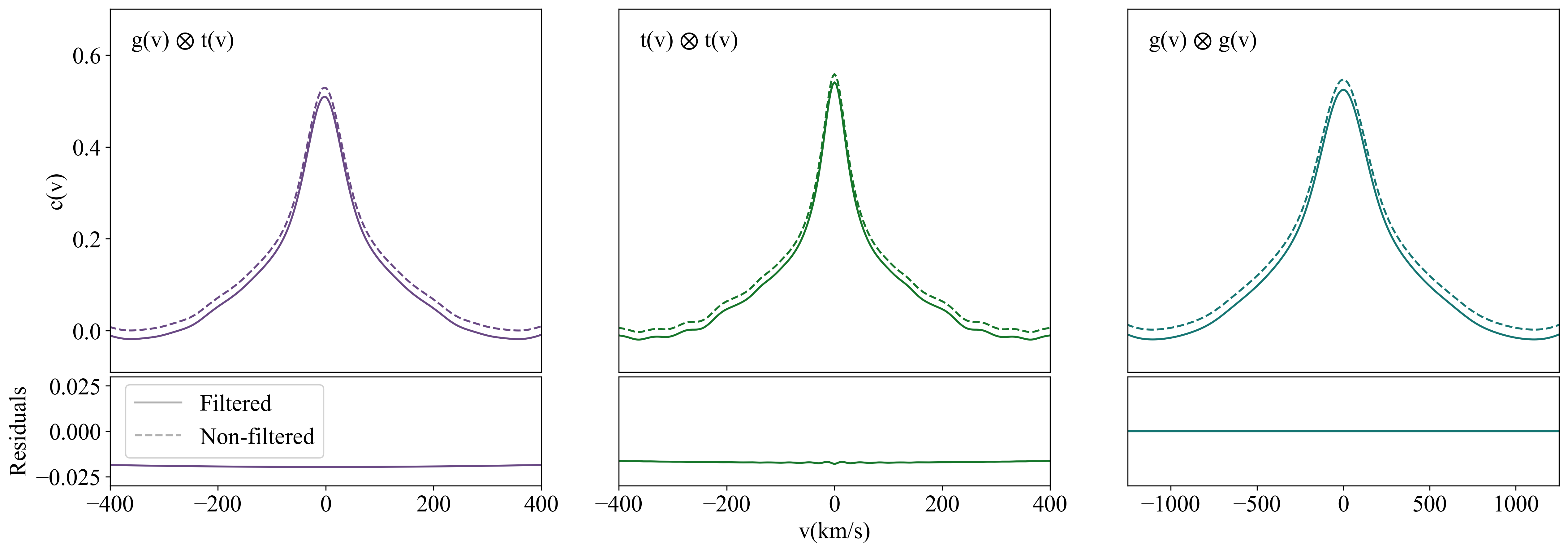}
\includegraphics[width=0.66\textwidth]{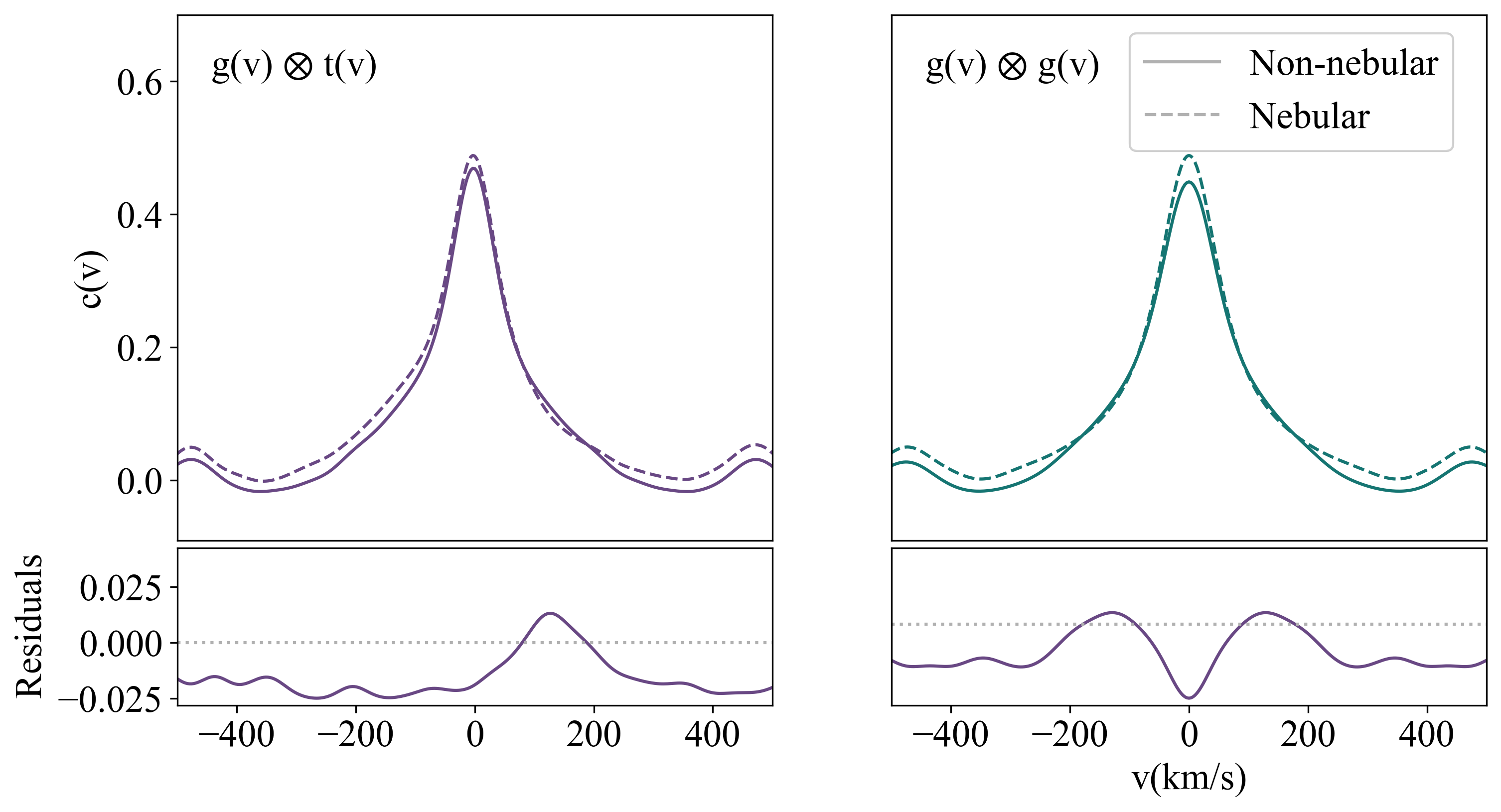}
\caption{upper panels: Cross correlations between frequency filtered and not filtered functions (solid and dotted lines respectively). From left to right are shown: g(v) $\otimes$ t(v), t(v) $\otimes$ t(v), and g(v) $\otimes$ g(v) in units of velocity. Lower panels:  Cross correlations between functions with and without nebular component (dotted and solid lines respectively). Left and right panels show the g(v) $\otimes$ t(v) and  g(v) $\otimes$ g(v) correlation functions.}
\label{fig:pruebas_mask_filt}
\end{figure*}

We have analysed the errors that can occur in two of the steps of the application of the method. The first one is the frequency filtering, and it is shown in the upper panels of Figure \ref{fig:pruebas_mask_filt} that show from left to right the galaxy-to-star, star-to-star and galaxy-to-galaxy cross correlation functions for MEGARA data. Two different lines, solid and dotted, show the results for a filtered spectrum using the values given above (see Sec. \ref{sec:megara_apli}) and a non-filtered one respectively. We can see that the filtering only has an effect on the step level of the correlation peak, hence the velocity dispersion calculations are not affected by errors associated with this part of the procedure.

The second important step is the subtraction of the non-stellar component from the spectrum. If we want to apply this method to galaxies or star clusters the removal of the nebular component is necessary. In order to evaluate the involved errors, we have simulated a nebula ionised by a young star cluster synthesised using the PopStar code \citep{Popstar} with Salpeter's IMF \citep{Salpeter1955} IMF with lower and upper mass limits of 0.85 and 120 M$_\odot$ respectively. We have selected an age of 5 Myr, and constant electron density of 100 cm$^{-3}$. A cluster with 4 $\times$ 10$^{4}$ M$_\odot$ and a nebular and stellar velocity dispersion of 20 km/s has been calculated to represent the simulated clusters. From the H$\beta$ calculated emission line intensity, the equivalent Paschen lines have been obtained with the PyNeb tool \citep{pyneb} for n$_e$(cm$^{-3}$) = 10$^2$ and T(K) = 10$^4$ values using \citet{Storey1995} atomic data.

The lower panels of Figure \ref{fig:pruebas_mask_filt} show the galaxy-to-star and the galaxy-to-galaxy cross correlations respectively, and the correlation without and with nebular emission lines are shown by solid and dotted lines. We can see that the shape of the correlation function peaks and their width are modified, introducing errors in the obtained results. In this particular case, the different measured velocities differ in 4.8 km/s.

\citet{1979AJ.....84.1511T} proposed estimating the internal error of the method from the asymmetric part of the correlation peak, calculating its root mean square. However, after our analysis we can conclude that, at high spectral resolution, this value is not representative of the real measurement error since our main error sources come from the spectrum preparation and the measurement procedure, and not from the quality of the data themselves. Thus, we propose estimating the velocity dispersion errors considering the asymmetries in the correlation peak caused by the entire procedure, calculating the largest and smallest Gaussian width that can be fitted. The corresponding error will be the semi-difference between these two values.

\section{Conclusions}
In this work, we have applied Tonry and Davis' technique  \citep{1979AJ.....84.1511T} to MUSE and MEGARA data in order to evaluate its validity at intermediate and high spectral resolutions. For this purpose, we have used as galaxy spectra convolutions of the stellar template spectrum with known Gaussian functions of different widths. The comparison between the input and output broadening functions shows that the application of Tonry and Davis' method is entirely compatible within the errors for MUSE data but this is not the case for the higher resolution data obtained by MEGARA. This is due to the main assumption of the cross-correlation technique not being fulfilled: the cross-correlation function does not present a Gaussian behavior in the star-to-star correlation at this high spectral resolution. 

Therefore, we have adapted Tonry and Davis' method by developing a mathematical equivalent one using the galaxy-to-galaxy correlation instead of the star-to-star correlation which is the one carrying the instrumental resolution information and therefore is the narrower of the three, hence in very high resolution data can become too narrow to be adequately represented by a Gaussian function. At this spectral resolution Tonry \& Davis' method fails entirely, overestimating the derived velocity dispersions by factors up to about 3.5. Taking into account that derived dynamical masses scale with the square of the velocity dispersion, these latter ones would be overestimated by up to a factor of about 10. According to these results the recommended method to be applied for high spectral resolution data is the one using the galaxy-to-galaxy and star-to-galaxy correlations. Results found by the usual application of Tonry \& Davis' method cannot be considered as correct when applied to high spectral resolution data. 

An error analysis shows that the internal errors of the method are unrealistically low and not representative of the real measurement errors. Therefore, we propose estimating errors by calculating the largest and smallest Gaussian widths that can be fitted to the principal correlation peak and using the semi-difference between these two values.

\section*{Acknowledgements}
This work has been supported by Spanish grants from the former Ministry of Economy, Industry and Competitiveness through the MINECO-FEDER research grant AYA2016-79724-C4-1-P, the present Ministry of Science and Innovation through research grant PID2019-107408GB-C42 and the National Research Agency through research grant AEI/10.13039/501100011033.

S.Z. acknowledges the support from contract: BES-2017-080509 associated to the first of these grants. 

\section*{Data Availability}
The data on which this article is based can be found in the web page of the MEGASTAR stellar library\footnote{https://www.fractal-es.com/megaragtc-stellarlibrary/private/home}.



\bibliographystyle{mnras}
\bibliography{Article} 





\bsp	
\label{lastpage}
\end{document}